\documentclass[prd,onecolumn]{revtex4}%
\usepackage{epsfig}
\usepackage{amsmath}
\usepackage{verbatim}
\usepackage{graphicx}%
\usepackage{amsfonts}%
\usepackage{amssymb}
\def\slash#1{\setbox0=\hbox{$#1$}                  \dimen0=\wd0                                    
\setbox1=\hbox{/} \dimen1=\wd1                  \ifdim\dimen0>\dimen1                              
\rlap{\hbox to \dimen0{\hfil/\hfil}}            #1                                           
\else                                              \rlap{\hbox to \dimen1{\hfil$#1$\hfil}}         /                                            \fi}

\begin{document}

\title{Infrared degrees of freedom of Yang-Mills theory in the Schr\"{o}dinger representation}
\author{Hilmar Forkel}
\affiliation{IFT, Universidade Estadual Paulista, Rua Pamplona, 145, 01405-900 S\~{a}o
Paulo, SP, Brazil }
\affiliation{ECT*, Strada delle Tabarelle 286, I-38050 Villazzano (Trento), Italy}
\affiliation{Institut f\"{u}r Theoretische Physik, Universit\"{a}t Heidelberg, D-69120
Heidelberg, Germany }

\begin{abstract}
We set up a new calculational framework for the Yang-Mills vacuum transition
amplitude in the Schr\"{o}dinger representation. After integrating out
hard-mode contributions perturbatively, we perform a gauge invariant gradient
expansion of the ensuing soft mode action which renders a subsequent saddle
point expansion for the vacuum overlap manageable. The standard ``squeezed''
approximation for the vacuum wave functional then allows for an essentially
analytical treatment of physical amplitudes. Moreover, it leads to the
identification of dominant and gauge invariant classes of gauge field orbits
which play the role of gluonic infrared (IR) degrees of freedom. Those emerge
as a rich variety of (mostly\ solitonic) solutions to the saddle point
equations which are characterized by a common relative gauge orientation of
the underlying gluon fields. We discuss their scale stability, guaranteed by a
virial theorem, and other general properties including topological quantum
numbers and action bounds. We then find important saddle point solutions
explicitly and examine their physical impact. Some of them are related to
tunneling solutions of the classical Yang-Mills equation, i.e. to instantons
and merons, while others appear to play unprecedented roles. A remarkable new
class of IR degrees of freedom comprises vortex and knot solutions of
Faddeev-Niemi type, potentially related to glueballs.

\end{abstract}
\maketitle
\preprint{IFT-P.033/2005}

\section{Introduction}

The strong couplings among soft QCD gluons manifest themselves in a variety of
complex long-distance phenomena. Most of them are thoroughly entwined with the
vacuum state, as illustrated by such prominent examples as quark confinement,
spontaneous chiral symmetry breaking, vacuum tunneling processes, the ensuing
$\theta$ structure as well as large gluon condensates. Despite the apparent
diversity of these and other effects, however, the essence of the underlying
dynamics is often expected to involve just a few soft gluonic modes.

The quest for these long-wavelength excitations began shortly after the
inception of QCD and has inspired the development of various kinds of vacuum
models, based e.g. on glueball condensation \cite{han82}, Gaussian stochastic
processes \cite{dos87}, gluonic domains \cite{nie79,kal01} and instanton
\cite{sch98} as well as meron \cite{dea76} ensembles. Over the last decade,
lattice simulations increasingly assisted in the search for predominant
infrared (IR) gluon fields, mostly by means of numerical ``filtering''
\cite{hoe87} and gauge-fixing \cite{che97} techniques. These simulations are
now beginning to generate quantitative insights into the role of instantons
and their size distribution \cite{instlat}, and into the classic confinement
scenarios based on (gauge-projected abelian) monopole \cite{che97,tho76} or
center vortex \cite{tho78,gre03} condensation.

Nevertheless, at present no mechanism involving soft vacuum gluons can be
uniquely or systematically related to QCD, and many crucial questions
regarding the underlying fields, their stability, gauge-independent physical
interpretation, mutual interactions, relations to other vacuum fields etc.
remain unanswered. Analytical progress in this realm has been particularly
slowed by the inevitable gauge dependence of the generally rather complex
classical gluon field configurations on which most of the existing proposals
are based. Similarly, approaches which reformulate non-Abelian gauge theory in
terms of gauge-independent loop variables \cite{mak81} or resolve the gauge
constraints explicitly (e.g. in Coulomb gauge \cite{chr80}), are often
technically too involved for direct practical applications.

In the present paper, we circumvent such complications by developing an
approach in which not the contributions of single gauge fields but rather
those of gauge invariant classes are treated jointly. These classes gather
contributions from dominant gauge field orbits to low-energy Yang-Mills
amplitudes and thus represent collective gluonic IR degrees of freedom.
Technically, they are the saddle points of a soft mode action for gauge
invariant matrix fields and therefore also provide the principal input for a
systematic saddle point expansion of soft Yang-Mills amplitudes.

Manifest gauge invariance is maintained throughout all calculations by working
in the Hamiltonian formulation of Yang-Mills theory in the ``coordinate''
Schr\"{o}dinger representation and by making use of explicit gauge projection
operators. The Schr\"{o}dinger picture is adopted mainly because it restricts
gauge transformations to a fixed reference time, thereby effectively
decoupling them from the dynamical time evolution, and because it often
renders the impact of topological gluon properties particularly transparent
(even without recourse to the semi-classical approximation) \cite{jac90}. For
the reasons already alluded to, we will focus on gluonic effects and work in
pure gauge theory without quarks.

The individual IR\ degrees of freedom, i.e. the solutions of the saddle point
equations, turn out to comprise a diverse range of specific features.
Additionally, they have several important properties in common, including
stability against scale transformations (an indispensable prerequisite for the
saddle point expansion which is ensured by a virial theorem) and
characteristic topological properties inherited from the gauge group. The
topology will turn out to be particularly useful for establishing relations
between specific saddle-point families and the instanton and meron solutions
of the classical Yang-Mills equation.

Our approach maintains explicitly traceable links between the soft
collective\ fields and the underlying gluon fields. The resulting IR dynamics
is at an intermediate level of complexity, somewhere inbetween the microscopic
theory itself and effective theories (e.g. for Polyakov loops) which just
share the symmetries of the fundamental dynamics while coupling parameters
have to be fitted to experimental (or lattice) data. Our soft-mode Lagrangian,
in contrast, follows uniquely from the adopted vacuum wave functional and
combines a reasonable amount of transparency with accessibility to essentially
analytical treatment.

The paper is organized as follows: in Sec. \ref{voa} we recapitulate the
definition of the vacuum overlap amplitude in the Schr\"{o}dinger picture. We
then implement a gauge-projected vacuum wave functional on the basis of the
Gaussian approximation and rewrite the overlap in terms of a bare action which
had previously emerged in a variational context. In Sec. \ref{sge}, we take
advantage of known 1-loop results to integrate out the hard-mode contributions
to the bare action perturbatively. By means of a controlled derivative
expansion, the renormalized soft-mode action density is then transformed it
into a local Lagrangian which lends itself to direct analytical treatment. In
Sec. \ref{spex} we build on these results by establishing the IR-sensitive
saddle point expansion for the functional integral over the soft modes and by
deriving the saddle point equations whose\ nontrivial\ solutions constitute
the new gluonic IR degrees of freedom.

Important generic properties of these IR variables are established in Sec.
\ref{sps}, including their scale stability due to a virial theorem, three
topological quantum numbers and a lower bound of Bogomol'nyi type on their
action. In Sec. \ref{solncl}, several classes of the more symmetric and most
important saddle point\ solutions are found explicitly. They comprise
topological soliton solutions of hedgehog type, which are related to classical
solutions of the Yang-Mills equation, and solutions which carry different
types of topological information and seem to have no obvious counterparts in
classical Yang-Mills theory. One of the most interesting solution classes
consists of solitonic links, twisted links and knots. Those emerge from a
generalization of Faddeev-Niemi theory which turns out to be embedded in our
soft-mode Lagrangian. In Sec. \ref{qspi} we classify all hedgehog soliton
solutions, find their most important representatives explicitly, and establish
the role of the regular solutions as mainly summarizing contributions from
instanton and meron gauge orbits to the vacuum overlap. In Sec. \ref{suc},
finally, we collect our principal results, comment on evaluating the
contributions of the gluonic IR degrees of freedom to relevant amplitudes, and
suggest directions for future work.

\section{Vacuum overlap amplitude}

\label{voa}

The vacuum overlap amplitude of $SU\left(  N\right)  $ Yang-Mills theory
(without matter fields)\ in the Schr\"{o}dinger ''coordinate'' representation
\cite{jac90} reads
\begin{equation}
Z^{\prime}:=\left\langle 0,t_{+}|0,t_{-}\right\rangle =\int D\vec{A}\Psi
_{0}^{\ast}\left[  \vec{A},t_{+}\right]  \Psi_{0}\left[  \vec{A},t_{-}\right]
.\label{zprime}%
\end{equation}
The vacuum wave functional (VWF) $\Psi_{0}$ depends on half of the canonical
variables, i.e. on \ the static gauge fields $\vec{A}\left(  \vec{x}\right)
$. Its gauge invariance, like that of any other physical state and wave
functional, is dictated by Gau\ss ' law. This crucial requirement can be
imposed on a given functional by simply projecting out its gauge-singlet
component, i.e. by integration over the (compact) gauge group \cite{pol78}.
Starting from an approximate and therefore generally gauge dependent wave
functional $\psi_{0}$, one then obtains the associated VWF
\begin{equation}
\Psi_{0}\left[  \vec{A}\right]  =\sum_{n}e^{iQ\theta}\int D\mu\left(
U^{\left(  Q\right)  }\right)  \psi_{0}\left[  \vec{A}^{U^{\left(  Q\right)
}}\right]  =:\int DU\psi_{0}\left[  \vec{A}^{U}\right]  \label{ginvvwf}%
\end{equation}
($d\mu$ is the invariant Haar measure of the gauge group, $Q$ is the homotopy
degree or winding number of the group element $U^{\left(  Q\right)  }$, and
$\theta$ is the vacuum angle) for which Gau\ss ' law is manifest. The vacuum
energy has been set to zero.

After interchanging the order of integration over gauge fields and gauge group
in Eq. (\ref{zprime}), it becomes obvious that a gauge group volume\ can be
factored out of $Z^{\prime}$, i.e.
\begin{equation}
Z^{\prime}=\int DU_{+}\int DU_{-}\int D\vec{A}\psi_{0}^{\ast}\left[  \vec
{A}^{U_{+}}\right]  \psi_{0}\left[  \vec{A}^{U_{-}}\right]  =:Z\int DU_{-},
\end{equation}
since the $\vec{A}$-integral is gauge invariant. In fact, the group volume
$Z^{\prime}/Z$ \ is left over when the two un-normalized gauge projectors in
the matrix element $Z^{\prime}$ are multiplied into one. The integrand of the
remaining integral over the gauge group is naturally rewritten as a Boltzmann
factor, i.e.
\begin{equation}
Z=\int DU\exp\left(  -\Gamma_{b}\left[  U\right]  \right)  ,\label{z}%
\end{equation}
which defines the 3-dimensional Euclidean bare action $\Gamma_{b}$ as a
functional of the ``relative'' gauge orientation $U\equiv U_{-}^{-1}U_{+}$
only. Owing to the gauge invariance of the gluon field measure, $\Gamma_{b}$
is gauge invariant as well and takes the explicit form
\begin{equation}
\Gamma_{b}\left[  U\right]  =-\ln\int D\vec{A}\psi_{0}^{\ast}\left[  \vec
{A}^{U}\right]  \psi_{0}\left[  \vec{A}\right]  .\label{gammab}%
\end{equation}
This action describes the dynamical correlations which the gauge projection of
the functional $\psi_{0}$ in Eq. (\ref{ginvvwf}) has generated. Hence it would
become trivial, i.e. $U$-independent, if $\psi_{0}$ were gauge invariant by
itself. More specifically, $\Gamma_{b}\left[  U\right]  $ gathers all those
contributions to $Z$ whose approximate vacua $\psi_{0}$ at $t=\pm\infty$
differ by the relative gauge orientation $U$. The variable $U$ thus represents
the contributions of a specifically weighted ensemble of all gluon field
orbits to the vacuum overlap and is gauge invariant by construction.

To proceed in an analytically tractable fashion, we now adopt the standard
Gaussian approximation
\begin{equation}
\psi_{0}^{\left(  G\right)  }\left[  \vec{A}\right]  =\exp\left[  -\frac{1}%
{2}\text{ }\int d^{3}x\int d^{3}yA_{i}^{a}\left(  \vec{x}\right)
G^{-1ab}\left(  \vec{x}-\vec{y}\right)  A_{i}^{b}\left(  \vec{y}\right)
\right]  \label{ga}%
\end{equation}
for the unprojected VWF \footnote{For our exploratory purposes we do not
retain a possible longitudinal contribution to $G^{-1}$ which was discussed in
Ref. \cite{dia98}. A minimal gauge-invariant extension of the exponent in Eq.
(\ref{ga}) characterizes the ground state of Yang-Mills theory in 2+1
dimensions \cite{kar98}.}, which has the decisive advantage of allowing
integrals over $\vec{A}$ to be done exactly. As expected from a ground state
wave functional, $\psi_{0}^{\left(  G\right)  }$ has no nodes. It describes a
``squeezed'' state, i.e. an oscillator-type extension of the unstable coherent
gluon states \cite{leu81} and thus the simplest natural candidate for the
vacuum functional. In fact, Eq. (\ref{ga}) turns into the exact ground state
for $U\left(  1\right)  $ gauge theory (up to color factors) if the
``covariance'' $G^{-1}$ is taken to be the inverse of the static vector
propagator. Several additional properties indicate that Gaussian VWFs with
suitably adapted covariances capture crucial features of the Yang-Mills
dynamics as well. Indeed, with an appropriate choice for $G^{-1}$ (see below)
the wave functional (\ref{ga}) becomes exact at high momenta and incorporates
asymptotic freedom. Moreover, it is known from variational analyses that
Gaussian VWFs generate a dynamical mass gap and possibly confinement
\cite{wan88,sze04}. (Mass generation and most other features of 2+1
dimensional compact photodynamics are also reproduced \cite{nol04}.)
Additional support for the Gaussian approximation will emerge from our results below.

After specializing the expression (\ref{gammab}) for the bare action to
$\psi_{0}^{\left(  G\right)  }$, the functional integral over the gluon fields
becomes Gaussian and can readily be carried out. The result takes the form of
a 3-dimensional, bilocal nonlinear sigma model \cite{kog95},
\begin{equation}
\Gamma_{b}\left[  U\right]  =\frac{1}{2g_{b}^{2}}\int d^{3}x\int d^{3}%
yL_{i}^{a}\left(  \vec{x}\right)  D^{ab}\left(  \vec{x}-\vec{y}\right)
L_{i}^{b}\left(  \vec{y}\right)  .\label{effact}%
\end{equation}
(Above we have omitted a term of higher order in the small bare coupling
$g_{b}$ which vanishes at the saddle points in which we will be interested
below.) The $U$-dependence enters $\Gamma_{b}$ both via the one-forms%
\begin{equation}
L_{i}\left(  \vec{x}\right)  =U^{\dagger}\left(  \vec{x}\right)  \partial
_{i}U\left(  \vec{x}\right)  =:L_{i}^{a}\left(  \vec{x}\right)  \frac{\tau
^{a}}{2i},\label{L}%
\end{equation}
i.e. the Lie-algebra valued, left-invariant Maurer-Cartan ``currents'' (with
real components $L_{i}^{a}$), and through higher-order\ corrections to the
bilocal operator
\begin{equation}
D^{ab}=\left[  \left(  G+G^{U}\right)  ^{-1}\right]  ^{ab}\simeq\frac{1}%
{2}G^{-1}\delta^{ab}+...
\end{equation}
where $G^{U}=G^{ab}\left(  \vec{x}-\vec{y}\right)  U^{\dagger}\left(  \vec
{x}\right)  T^{a}U\left(  \vec{x}\right)  \otimes U\left(  \vec{y}\right)
T^{b}U^{\dagger}\left(  \vec{y}\right)  $, $T^{a}=\lambda^{a}/2$ and
$G^{-1ab}=G^{-1}\delta^{ab}$. The above reformulation of the Yang-Mills vacuum
overlap on the basis of a gauge-invariant Gaussian VWF was employed in Ref.
\cite{kog95} as the starting point for a variational approach
\footnote{Several other variational and related Schwinger-Dyson equation
studies on the basis of the Gaussian trial state have been performed in
gauge-fixed formulations \cite{wan88,sze04}.}. Alternatively, it can be
obtained from a saddle point evaluation of the functional integral
\cite{dia98} in Eq. (\ref{gammab}) which becomes exact for the Gaussian VWF
\footnote{Strictly speaking, the gauge transformations of Yang-Mills theory
without matter span the coset $SU\left(  N\right)  /Z_{N}$ since center
elements of $SU\left(  N\right)  $ act trivially on the gauge fields. We
refrain from making this restriction manifest here, although it could be
implemented in the action (\ref{effact}) and may become relevant, e.g., for
the discussion of center vortices.}.

Although the nonlinear sigma model (\ref{effact}) is easier to handle than the
original Yang-Mills theory, its exact non-perturbative treatment remains
beyond analytical reach \cite{fer05}. Nevertheless, the parametric enhancement
of the action (\ref{effact}) by the factor $g_{b}^{-2}$ suggests that a useful
approximation may be obtained from a saddle point expansion of the functional
integral (\ref{z}). In order to render this approximation practical, however,
one has to deal with the nonlocality of the bare action (\ref{effact}) which
encumbers the identification and evaluation of the saddle points. We will show
in the following section that this can be efficiently accomplished by
combining a renormalization group evolution of the bare action (which removes
the explicit UV modes) with a subsequent derivative expansion to transform the
IR dynamics into an approximately local soft-mode action.

\section{Soft gradient expansion}

\label{sge}

For the reasons outlined in the introduction, we are mainly interested in soft
Yang-Mills amplitudes with external momenta $\left|  \vec{p}_{i}\right|  $
smaller than a typical hadronic scale $\mu$ (the lowest glueball mass, for
example). This restricted focus permits us to recast the bare action
(\ref{effact}) into a form which only retains soft field modes explicitly and
which can be systematically approximated by a local Lagrangian. The present
section describes the derivation and some useful features of this soft-mode action.

Since the action (\ref{effact}) incorporates asymptotic freedom (for proper
choices of $G$, see below), the bare coupling $g_{b}$ is small at the large
cutoff scale $\Lambda_{UV}$ where the theory is originally defined. The hard
modes of the $U$ field with momenta $\left|  \vec{k}\right|  >\mu$ can
therefore be integrated out of the functional integral (\ref{z})
perturbatively, down to values of the infrared scale $\mu$ where the
renormalized coupling $g\left(  \mu\right)  $ ceases to be much smaller than
unity. In practice, this may be done for instance by Wilson's momentum-shell
technique \cite{wil74}, after factorizing $U$ into contributions from high-
and low-frequency modes. To one-loop order, the resulting renormalization of
the action just amounts to the replacement of the bare coupling $g_{b}$ by the
running coupling $g\left(  \mu\right)  $. This was confirmed in Ref.
\cite{bro99} where the one-loop coupling was obtained as
\begin{equation}
g\left(  \mu\right)  =g_{b}+\frac{g_{b}^{3}N_{c}}{\left(  2\pi\right)  ^{2}%
}\ln\frac{\Lambda_{UV}}{\mu}+O\left(  g_{b}^{5}\right)  \label{g(mu)}%
\end{equation}
(for $G\left(  k\right)  =k^{-1}$ at $k>\mu$). The scaling behavior of
$g\left(  \mu\right)  $ makes asymptotic freedom explicit and reaffirms that
the Gaussian VWF reproduces the qualitative UV behavior of Yang-Mills theory.
In fact, Eq. (\ref{g(mu)}) equals the one-loop Yang-Mills\ coupling up to a
small correction factor $1/11$ which arises from the absence of transverse
gluons and could be avoided by introducing an anisotropic component for
$G^{-1}$ \cite{dia98}. The one-loop integration over the high-momentum modes
was found to be reliable down to $\mu\simeq1.3$ GeV\ \cite{bro99} which
provides a useful benchmark for numerical estimates. In the valid range of
$\mu$ values Eq. (\ref{effact}) turns into the renormalized soft-mode action
\begin{equation}
\Gamma\left[  U_{<}\right]  =\frac{1}{4g^{2}\left(  \mu\right)  }\int
d^{3}x\int d^{3}yL_{<,i}^{a}\left(  \vec{x}\right)  G^{-1}\left(  \vec{x}%
-\vec{y}\right)  L_{<,i}^{b}\left(  \vec{y}\right)  \label{gamless}%
\end{equation}
where the subscript $"<"$ indicates that $U$ contains only $k<\mu$ modes.

The action (\ref{gamless}) is still nonlocal. However, this nonlocality is
substantially weaker than in the bare action (\ref{effact}) since the soft $U$
fields in the integrand vary too little to resolve details of $G^{-1}$ over
distances smaller than $\mu^{-1}$. This observation can be turned into a
controlled, \emph{local} approximation scheme for the soft-mode action
(\ref{gamless}) by exploiting the fact that the gradients of $U_{<}$ are
bounded by the IR gluon mass scale,
\begin{equation}
\left|  \vec{\partial}U_{<}\right|  <\mu.
\end{equation}
Indeed, this bound suggests to expand the nonlocality of $G^{-1}$ into
derivatives $\partial/\mu$ which will act upon $U_{<}$ after partial
integration. Using the isotropy of $G_{ij}^{-1}=G^{-1}\delta_{ij}$ (as
mentioned above, an anisotropic component could be allowed in principle and
would lead to somewhat more general expressions), one has
\begin{equation}
G^{-1}\left(  \vec{x}-\vec{y}\right)  =\mu\left[  c_{0}+c_{1}\frac{\vec
{\partial}_{x}^{2}}{\mu^{2}}+c_{2}\left(  \frac{\vec{\partial}_{x}^{2}}%
{\mu^{2}}\right)  ^{2}+c_{3}\left(  \frac{\vec{\partial}_{x}^{2}}{\mu^{2}%
}\right)  ^{3}+...\right]  \delta^{3}\left(  \vec{x}-\vec{y}\right)  .
\label{gexp}%
\end{equation}
The dimensionless constants $c_{i}$ encode the low-momentum behavior of
$G^{-1}$ and could, e.g., be determined variationally. For our present
purposes, however, it will be sufficient to adopt the standard expression
$G^{-1}\left(  \vec{k}\right)  =\sqrt{\vec{k}^{2}+\mu^{2}}$ which approximates
the solution of Schwinger-Dyson equations and variational estimates
\cite{wan88,sze04} and incorporates both asymptotic freedom and a dynamical
mass gap. The corresponding $c_{i}$ can be read off\ directly from the Fourier
transform
\begin{equation}
G^{-1}\left(  \vec{x}-\vec{y}\right)  =\sqrt{-\vec{\partial}_{x}^{2}+\mu^{2}%
}\delta^{3}\left(  \vec{x}-\vec{y}\right)  =-\frac{1}{2\pi^{2}}\frac{\mu
^{2}K_{2}\left(  \mu\left|  x-y\right|  \right)  }{\left|  x-y\right|  ^{2}}
\label{gm1x}%
\end{equation}
($K_{2}$ is a McDonald function \cite{abr}), i.e. $c_{0}=1,$ $c_{1}=-1/2,$
$c_{2}=-1/8$ etc.

The combination of the above results leads to the intended reformulation of
the nonlocal dynamics (\ref{effact}). As anticipated, the bilocal action
density for the soft modes in Eq. (\ref{gamless}) becomes a (``quasi''-) local
Lagrangian $\mathcal{L}\left(  \vec{x}\right)  $ and the action takes the
familiar form
\begin{equation}
\Gamma\left[  U_{<}\right]  =\int d^{3}x\mathcal{L}\left(  \vec{x}\right)
.\label{gamsoft}%
\end{equation}
The Lagrangian is an expansion into powers of $\mu^{-1}\partial_{i}U_{<}$ and
therefore belongs to the class of generalized nonlinear sigma models. When
expressed in terms of the Cartan-Maurer currents $L_{<,i}$, it reads
\footnote{The Lagrangian (\ref{efflagr}) could immediately be generalized to
arbitrary $G^{-1}$ by restoring the original $c_{i}$ dependence from Eq.
(\ref{gexp}).}%
\begin{equation}
\mathcal{L}\left(  \vec{x}\right)  =-\frac{\mu}{2g^{2}\left(  \mu\right)
}tr\left\{  L_{<,i}\left(  \vec{x}\right)  L_{<,i}\left(  \vec{x}\right)
+\frac{1}{2\mu^{2}}\partial_{i}L_{<,j}\left(  \vec{x}\right)  \partial
_{i}L_{<,j}\left(  \vec{x}\right)  -\frac{1}{8\mu^{4}}\partial^{2}%
L_{<,i}\left(  \vec{x}\right)  \partial^{2}L_{<,i}\left(  \vec{x}\right)
+...\right\}  .\label{efflagr}%
\end{equation}
We have omitted total derivatives $\Delta\mathcal{L}$ from the higher-order
terms of $\mathcal{L}$ since they do not affect the field equations.
Nevertheless, they may generate non-vanishing surface terms due to
infinite-action configurations which are generally irrelevant for the saddle
point expansion. To lowest order,%
\begin{equation}
\Delta\mathcal{L}\left(  \vec{x}\right)  =\frac{1}{8g^{2}\left(  \mu\right)
\mu}tr\left\{  \partial^{2}\left[  L_{<,i}\left(  \vec{x}\right)
L_{<,i}\left(  \vec{x}\right)  \right]  +...\right\}  .\label{sfterm}%
\end{equation}

The gradient expansion in Eq. (\ref{efflagr}) is controlled by increasing
powers of the parametrically small $\left(  \partial U_{<}/\mu\right)  ^{2}$.
For practical purposes it can therefore be reliably truncated, at an order
which is determined by the desired accuracy of the approximation to the exact
action. Below, we will be interested in specific field configurations (saddle
points) which we are going to find explicitly. For those one may directly
check \emph{a posteriori} whether the full action is sufficiently well
reproduced and, if not, include contributions from higher gradients. Based on
such tests and a virial theorem (see Sec. \ref{virth}), we found the
truncation at $O\left(  \left(  \partial/\mu\right)  ^{4}\right)  $ to yield a
generally sufficient approximation (at the few percent level) to the full
action (\ref{gamless}).

The first term\ in the Lagrangian (\ref{efflagr})\ has the standard form of a
3-dimensional nonlinear $\sigma$-model (or principal chiral model). The second
one, with four derivatives acting on $U_{<}$'s, is reminiscent of a similar
term in the Skyrme model \cite{zah86}. However, the identity%
\begin{equation}
\partial_{i}L_{j}\partial_{i}L_{j}=\frac{1}{2}\left[  L_{i},L_{j}\right]
^{2}+\partial_{i}L_{j}\partial_{j}L_{i}\label{commplus}%
\end{equation}
shows that the four-derivative contribution to Eq. (\ref{efflagr}) contains,
besides the commutator or Skyrme term, a part which qualitatively alters the
character of the Euler-Lagrange equations. While the commutator generates only
second-order terms to the field equations (see below), the piece without
equivalent in the Skyrme model leads to additional fourth-order terms which
allow for new types of solutions. (Several families of topological soliton
solutions from the Lagrangian (\ref{efflagr}) will nevertheless turn out to
resemble static Skyrmions \footnote{It should be kept in mind, of course, that
our $U$-field arises from Yang-Mills theory even without quarks, and that its
physical interpretation (representing gauge fields of a fixed relative color
orientation) is completely different from that in the Skyrme model where $U$
represents fluctuations around the chiral order parameter.}.)

We end this section by emphasizing that the construction of an analogous
gradient expansion in terms of the original gauge fields $\vec{A}$ would
require the (residual) gauge freedom to be completely fixed and thus give rise
to all the associated conceptual and calculational complications. (Otherwise a
``soft'' gauge field could just be turned into a\ ``hard'' one by a suitable
gauge transformation and would spoil the\ ``convergence'' of the derivative
expansion.) The locality and structural simplicity of the soft-mode Lagrangian
(\ref{efflagr}) can therefore be regarded as a benefit of reformulating the
dynamics in terms of the gauge invariant $U$ field variables.

\section{Saddle point expansion}

\label{spex}

Our next task will be to set up the saddle point (or, more specifically,
steepest descent)\ expansion of the functional integral over the soft modes,
\begin{equation}
Z=\int DU_{<}\exp\left(  -\Gamma\left[  U_{<}\right]  \right)  ,\label{zsoft}%
\end{equation}
where the vacuum overlap $Z$ serves as the prototype for similar integrals in
other soft amplitudes.\ This expansion is based on the IR\ saddle point fields
$\bar{U}_{i}\left(  \vec{x}\right)  $, i.e. the local minima of the soft-mode
action (\ref{gamsoft}). The search for these minima is simplified by the fact
that all finite-action $U$-field configurations, including most saddle points,
fall into disjoint topological classes (cf. Sec. \ref{top} and below). Since
fields which carry different topological charges - for now summarily denoted
by $Q$ - cannot be continuously deformed into each other, the local variation
of the action may be performed in each topological sector separately. This
amounts to solving the saddle point equations%
\begin{equation}
\left.  \frac{\delta\Gamma\left[  U_{<}\right]  }{\delta U_{<}\left(  \vec
{x}\right)  }\right|  _{U_{<}=\bar{U}_{i}^{\left(  Q\right)  }}=0\label{spaeq}%
\end{equation}
at fixed $Q$.

To leading order, the saddle-point expansion for the vacuum overlap $Z$ can
then be assembled by summing the contributions from the solutions
$\bar{U}_{i}^{\left(  Q\right)  }$ of Eq. (\ref{spaeq}), i.e.
\begin{equation}
Z\simeq\sum_{Q\in Z,i}F_{i}\left[  \bar{U}_{i}^{\left(  Q\right)  }\right]
\exp\left(  -\Gamma\left[  \bar{U}_{i}^{\left(  Q\right)  }\right]  \right)
,\label{zspa}%
\end{equation}
where the pre-exponential factors $F_{i}$ are generated by zero-mode
contributions which typically arise if continuous symmetries of the action are
broken by the solutions. The sum over the saddle points, labeled by $i$,
symbolically includes integrals with the appropriate measure when the saddle
points come in continuous families. Functional integrals for more complex
amplitudes, including the gluonic Green functions, receive contributions from
the same saddle points and are obtained by differentiating $Z$ with respect to
suitably implemented sources. Higher-order corrections to the approximation
(\ref{zspa}) can be systematically calculated from the (nondegenerate)
fluctuations around the solutions $\bar{U}_{i}^{\left(  Q\right)  }$. The
reliability of the leading-order approximation (\ref{zspa}) increases with the
action values of the saddle points because $\Gamma\left[  \bar{U}\right]
\gg1$ prevents the saddle point contributions from being rendered
insignificant by the fluctuations of $O\left(  1\right)  $ around them. In our
case, compliance with this criterion is reinforced by the overall factor
$g^{-2}\left(  \mu\right)  \gg1$ (for $\mu\gtrsim1.3-1.5$ GeV) in the
Lagrangian (\ref{efflagr}) which parametrically enhances the action
\footnote{The enhancement of the soft-mode action is somewhat weaker than that
of the bare action (\ref{effact}) since $g\left(  \mu\right)  \geq g_{b}$.}.

For the explicit solution of the saddle-point equation (\ref{spaeq}), as well
as for part of our general analysis below, it is practically necessary to
adopt a parametrization of the $SU\left(  N\right)  $ group elements $U$ which
allows to work directly with their $N^{2}-1$ independent degrees of freedom.
We will use the exponential representation
\begin{equation}
U_{<}\left(  \vec{x}\right)  =\exp\left[  \phi\left(  \vec{x}\right)  \hat
{n}^{a}\left(  \vec{x}\right)  \frac{\tau^{a}}{2i}\right]  \label{uparam}%
\end{equation}
for this purpose, where the coefficient vector of the Lie algebra generators
is decomposed into its direction, specified by the vector field $\hat{n}^{a}$
with $\hat{n}^{a}\hat{n}^{a}=1$ (which parametrizes the coset $SU\left(
N\right)  /H$ where $H$ is the Cartan subgroup of the gauge group), and its
length\textbf{ }$\phi$. For simplicity, we will also specialize our following
discussion to $N=2$. The soft-mode Lagrangian (\ref{efflagr}) can then be
rewritten in terms of the unit vector field $\hat{n}^{a}$ and the spin-0 field
$\phi$ as
\begin{equation}
\mathcal{L}\left(  U\right)  =\mathcal{L}_{2d}\left(  \phi,\hat{n}\right)
+\mathcal{L}_{4d}\left(  \phi,\hat{n}\right)  \label{l24d}%
\end{equation}
where, for the reasons discussed in the paragraph below Eq. (\ref{sfterm}),
all terms of the gradient expansion with up to four derivatives on the
$U$\ fields are retained. The two-derivative part $\mathcal{L}_{2d}$, i.e. the
standard nonlinear $\sigma$-model, becomes%
\begin{equation}
\mathcal{L}_{2d}=\frac{\mu}{2^{2}g^{2}\left(  \mu\right)  }\left[  \left(
\partial_{i}\phi\right)  ^{2}+2\left(  1-\cos\phi\right)  \left(  \partial
_{i}\hat{n}^{a}\right)  ^{2}\right]
\end{equation}
while the four-derivative contributions turn into%
\begin{align}
\mathcal{L}_{4d} &  =\frac{1}{2^{3}g^{2}\left(  \mu\right)  \mu}\left[
\left(  \partial^{2}\phi-\sin\phi\left(  \partial_{i}\hat{n}^{a}\right)
^{2}\right)  ^{2}+2\left(  1+\cos\phi\right)  \left(  \partial_{i}\phi
\partial_{i}\hat{n}^{a}\right)  ^{2}\right.  +2\left(  1-\cos\phi\right)
\left(  \varepsilon_{ijk}\partial_{j}\phi\partial_{k}\hat{n}^{a}\right)
^{2}\nonumber\\
&  +4\sin\phi\partial_{j}\phi\partial_{j}\hat{n}^{a}\partial^{2}\hat{n}%
^{a}+2\left(  1-\cos\phi\right)  \left(  \varepsilon^{abc}\hat{n}^{b}%
\partial^{2}\hat{n}^{c}\right)  ^{2}+\left.  2\left(  1-\cos\phi\right)
^{2}\left(  \varepsilon^{abc}\partial_{i}\hat{n}^{b}\partial_{j}\hat{n}%
^{c}\right)  ^{2}\right]  .
\end{align}
The general expressions above show that $\mathcal{L}\geq0$, as required for
the existence of the functional integral (\ref{zsoft}). The same remains true
if higher orders of the derivative expansion (\ref{efflagr}) are included (as
long as $\partial U_{<}/\mu<1$, of course). For the analysis of some generic
saddle point solution properties, including their behavior under scale
transformations (cf. Sec. \ref{sps}), it will prove useful that both
$\mathcal{L}_{2d}$ and $\mathcal{L}_{4d}$ are even individually nonnegative.

By varying\ the action (\ref{gamsoft}) of the Lagrangian (\ref{l24d}) with
respect to $\phi$ and $\hat{n}^{a}$, the saddle point equation (\ref{spaeq})
turns into a nonlinear system of four coupled partial differential equations.
For $\phi$ one directly obtains%
\begin{align}
\left(  4g^{2}\mu\right)  \frac{\delta\Gamma\left[  \phi,\hat{n}\right]
}{\delta\phi\left(  \vec{x}\right)  } &  =\partial^{4}\phi-2\partial^{2}%
\phi\left(  \partial_{i}\hat{n}^{a}\right)  ^{2}-4\partial_{i}\phi\partial
_{i}\partial_{k}\hat{n}^{a}\partial_{k}\hat{n}^{a}-4\cos\phi\left(
\partial_{i}\partial_{j}\phi\partial_{i}\hat{n}^{a}+\partial_{j}\phi
\partial^{2}\hat{n}^{a}+\partial_{i}\phi\partial_{i}\partial_{j}\hat{n}%
^{a}\right)  \partial_{j}\hat{n}^{a}\nonumber\\
&  +2\sin\phi\left[  \left(  \partial_{i}\phi\partial_{i}\hat{n}^{a}\right)
^{2}-\left(  1-\cos\phi\right)  \left(  \partial_{i}\hat{n}^{a}\partial
_{j}\hat{n}^{a}\right)  ^{2}-2\partial_{i}\hat{n}^{a}\partial_{i}\partial
^{2}\hat{n}^{a}-\left(  \partial_{i}\partial_{j}\hat{n}^{a}\right)
^{2}+\left(  \partial_{i}\hat{n}^{a}\right)  ^{2}\left(  \partial_{j}\hat
{n}^{b}\right)  ^{2}\right]  \nonumber\\
&  -\sin\phi\cos\phi\left(  \partial_{i}\hat{n}^{a}\right)  ^{2}\left(
\partial_{j}\hat{n}^{b}\right)  ^{2}-\sin\phi\left(  \delta^{ab}+\hat{n}%
^{a}\hat{n}^{b}\right)  \partial^{2}\hat{n}^{a}\partial^{2}\hat{n}^{b}%
-2\mu^{2}\left[  \partial^{2}\phi-\sin\phi\left(  \partial_{i}\hat{n}%
^{a}\right)  ^{2}\right]  =0.\label{omeq}%
\end{align}
The three equations for the components of the $\hat{n}$ field, on the other
hand, have to be derived by a constrained variation whose task it is to
preserve the unit length of $\hat{n}$. As a consequence, they can be cast into
the succinct form
\begin{equation}
\left(  \delta^{ab}-\hat{n}^{a}\hat{n}^{b}\right)  \frac{\delta\Gamma\left[
\phi,\hat{n}\right]  }{\delta\hat{n}^{a}\left(  \vec{x}\right)  }=0\label{neq}%
\end{equation}
where the projection operator ensures that the action is only affected by
those variations $\delta\hat{n}^{a}$ which maintain orthogonality to $\hat
{n}^{a}$. The evaluation of the functional derivative with respect to $\hat
{n}$ yields
\begin{align}
\left(  2g^{2}\mu\right)  \frac{\delta\Gamma\left[  \phi,\hat{n}\right]
}{\delta\hat{n}^{a}\left(  \vec{x}\right)  } &  =-\left(  1-\cos\phi\right)
\left(  \partial^{2}\hat{n}^{a}\hat{n}^{c}\partial^{2}\hat{n}^{c}\right)
+\partial_{i}\left\{  \sin\phi\partial^{2}\phi\partial_{i}\hat{n}^{a}\right.
-\sin\phi\partial_{i}\phi\partial^{2}\hat{n}^{a}-2\cos\phi\partial_{i}%
\phi\partial_{j}\phi\partial_{j}\hat{n}^{a}\nonumber\\
&  +\cos\phi\partial_{i}\phi\partial_{k}\phi\partial_{k}\hat{n}^{a}+\sin
\phi\partial_{i}\partial_{k}\phi\partial_{k}\hat{n}^{a}+\sin\phi\partial
_{k}\phi\partial_{i}\partial_{k}\hat{n}^{a}+\sin\phi\partial_{i}\phi\left(
\partial^{2}\hat{n}^{a}-\hat{n}^{a}\hat{n}^{c}\partial^{2}\hat{n}^{c}\right)
\nonumber\\
&  +\left(  1-\cos\phi\right)  \left[  \partial_{i}\partial^{2}\hat{n}%
^{a}-\hat{n}^{c}\partial_{i}\hat{n}^{a}\partial^{2}\hat{n}^{c}-\hat{n}%
^{a}\partial_{i}\hat{n}^{c}\partial^{2}\hat{n}^{c}-\hat{n}^{a}\hat{n}%
^{c}\partial_{i}\partial^{2}\hat{n}^{c}-\left(  \partial_{j}\phi\right)
^{2}\partial_{i}\hat{n}^{a}-2\mu^{2}\partial_{i}\hat{n}^{a}\right]
\nonumber\\
&  -\sin^{2}\phi\partial_{i}\hat{n}^{a}\left(  \partial_{i}\hat{n}^{c}\right)
^{2}-2\left(  1-\cos\phi\right)  ^{2}\left[  \partial_{i}\hat{n}^{a}\left(
\partial_{j}\hat{n}^{c}\right)  ^{2}-\partial_{i}\hat{n}^{c}\partial_{j}%
\hat{n}^{c}\partial_{j}\hat{n}^{a}\right]  \left.  {}\right\}  .
\end{align}
The saddle point equations (\ref{omeq}) and (\ref{neq}) are independent of the
coupling $g$ because it enters the action only through the overall factor
$g^{-2}$. In general, their solutions have to be found numerically. However,
we will demonstrate below that several nontrivial analytical solutions exist
and that further important solution classes with a rather high degree of
symmetry can be obtained by solving substantially simplified field equations.

Moreover, essential qualitative solution properties can be derived without
solving the saddle point equations explicitly (cf. Sec. \ref{sps}). Each
topological charge sector contains at least one action minimum, for example,
i.e. one solution of Eqs. (\ref{omeq}) and (\ref{neq}). The requirement of
finding and including all of them would render the saddle point expansion
practically useless. Fortunately, however, this turns out to be unnecessary.
Below we will establish lower bounds on the action of the saddle points which
are monotonically increasing functions of their topological charges. Hence
contributions from saddle points in high topological charge sectors can
generally be ignored.

\section{General properties of the saddle point solutions}

\label{sps}

Before actually solving the saddle-point equations (\ref{omeq}) and
(\ref{neq}) in Secs. \ref{solncl} and \ref{qspi}, it will be useful to obtain
a few general insights into the topological structure and stability properties
of the solutions. This is the objective of the present section.

\subsection{Virial theorem}

\label{virth}

In order to analyze the scaling behavior of the extended saddle point
solutions and to establish the underlying virial theorem, we define the scaled
fields
\begin{equation}
\phi_{\lambda}:=\phi\left(  \lambda\vec{x}\right)  ,\text{ \ \ \ \ }\hat
{n}_{\lambda}^{a}:=\hat{n}^{a}\left(  \lambda\vec{x}\right)
\end{equation}
for real $\lambda\neq0$ and note that $\mu$ is the only mass scale in the
field equations (\ref{omeq}) and (\ref{neq}). This immediately implies that
the solutions of the saddle point equations with scaled parameter
$\mu\rightarrow\lambda^{-1}\mu$ can be obtained by rescaling the original
solutions $\left(  \bar{\phi},\hat{n}\right)  $ to $\left(  \bar{\phi
}_{\lambda},\hat{n}_{\lambda}\right)  $.

In the following, however, we will keep $\mu$ fixed. The scale-transformed
extended solutions then cease to solve the field equations, and this simple
observation together with two basic properties of the Lagrangian (\ref{l24d})
can be turned into a virial theorem. The first step towards its derivation
consists in establishing the relation between the actions of scaled and
unscaled fields (as long as they stay finite), which can be read off from the
2- and 4-derivative parts of the Lagrangian (\ref{l24d}) separately:
\begin{equation}
\Gamma\left(  \lambda\right)  :=\Gamma\left[  \phi_{\lambda},\hat{n}_{\lambda
}\right]  =\Gamma_{2d}\left(  \lambda\right)  +\Gamma_{4d}\left(
\lambda\right)  =\frac{1}{\lambda}\Gamma_{2d}\left(  1\right)  +\lambda
\Gamma_{4d}\left(  1\right)  .\label{gamlam}%
\end{equation}
The second relevant property of the action based on Eq. (\ref{l24d}) is its
strict positivity for extended, i.e. not translationally invariant field
configurations (cf. Sec. \ref{spex}),
\begin{equation}
\Gamma_{2d}\left(  1\right)  ,\text{ }\Gamma_{4d}\left(  1\right)  >0.
\end{equation}
The remaining step is to specialize the fields under consideration to the
saddle point solutions. Since those extremize the action under arbitrary small
variations - which of course include infinitesimal scale transformations - one
immediately has $\left.  d\Gamma\left(  \lambda\right)  /d\lambda\right|
_{\lambda=1}=0$ and consequently the virial theorem
\begin{equation}
\Gamma_{2d}\left(  1\right)  =\Gamma_{4d}\left(  1\right)  .\label{vth}%
\end{equation}
Eq. (\ref{vth}) clearly exhibits the crucial role of the four-derivative terms
$\Gamma_{4d}$ in guaranteeing the stability of the saddle point solutions: for
$\Gamma_{4d}=0$ the remaining nonlinear $\sigma$-model action would be
minimized by sending $\lambda\rightarrow\infty$ (cf. Eq. (\ref{gamlam})), i.e.
by the scale collapse of the solutions which Derrick's theorem predicts
\cite{der64}. Our\ truncation of the gradient expansion (\ref{efflagr}) at
$O\left(  \left(  \partial U_{<}/\mu\right)  ^{2}\right)  $ therefore turns
out to be the minimal approximation which can support localized, stable
soliton solutions \footnote{The signs of the higher-order terms in the
derivative expansion (\ref{efflagr}) alternate. One might therefore suspect
that their contributions could destabilize the extended solutions. However,
this would just indicate an incorrect truncation of the gradient expansion.
Indeed, if a localized solution is small enough to be significantly affected
by higher-derivative terms, those would have to be added to the field
equations in the first place. (A somewhat analogous situation is encountered
in one-loop Coleman-Weinberg potentials: their physically relevant local
minima are stable only as long as fluctuations remain small, i.e. as long as
the underlying truncation of the loop expansion is justified \cite{col73}.)}.
Furthermore,
\begin{equation}
\left.  \frac{d^{2}\Gamma\left(  \lambda\right)  }{d\lambda^{2}}\right|
_{\lambda=1}=2\Gamma_{2d}\left(  1\right)  \geq0
\end{equation}
implies that the scaling extrema are indeed action minima and that solutions
with $\Gamma_{2d}\left(  1\right)  =0$ are points of inflection.

Finally, it is worth emphasizing that the coexistence of terms with different
numbers of derivatives in the Lagrangian (\ref{efflagr}), and thus ultimately
the virial theorem (\ref{vth}) and the existence of stable solutions, is
brought about by the mass scale $\mu$ which reflects the short-wavelength
quantum fluctuations integrated out in Sec. \ref{sge}. This situation is
reminiscent of the classical instanton solutions to the Yang-Mills equation
whose typical size scale must likewise be generated by quantum fluctuations.

To summarize, we have established a virial theorem which ensures that the
extended solutions of Eqs. (\ref{omeq}) and (\ref{neq}) are stable against
scale transformations. In our context this is an indispensable property since
unstable solutions would prohibit a useful saddle point expansion. As a side
benefit, the virial theorem also\ provides stringent tests for the numerical
solutions of the saddle point equations.

\subsection{Topology}

\label{top}

As a three-dimensional principal chiral model with stabilizing
higher-derivative terms, the soft-mode Lagrangian (\ref{l24d}) allows for
topological soliton solutions. In the present section we discuss three
topological invariants or charges which such solutions and more general
continuous fields $U$ may carry \cite{man04}.

The most fundamental topological classification arises from the fact that all
$U$-fields with a finite action $\Gamma$ based on any truncation of the
Lagrangian (\ref{efflagr}) have to approach the same constant at $\left|
\vec{x}\right|  \rightarrow\infty$. As a higher-dimensional analog of the
stereographic projection, this compactifies their domain to a three-sphere
$S_{sp}^{3}$. All $U$'s with finite $\Gamma$ therefore describe continuous
maps from $S_{sp}^{3}$ into the ``topologically active'' part of the group
manifold. For $SU\left(  N\right)  $ the latter is the trivially embedded
subgroup $SU\left(  2\right)  \sim S_{G}^{3}$. The resulting maps $S_{sp}%
^{3}\rightarrow S_{G}^{3}$ fall into disjoint homotopy classes, the elements
of the third homotopy group $\pi_{3}\left(  S^{3}\right)  =Z$, which are
characterized by a topological degree or charge%
\begin{equation}
Q\left[  U\right]  =\frac{1}{24\pi^{2}}\int d^{3}x\varepsilon_{ijk}tr\left\{
U^{\dagger}\partial_{i}UU^{\dagger}\partial_{j}UU^{\dagger}\partial
_{k}U\right\}  .\label{q}%
\end{equation}
(The integrand can be shown to be a total derivative, as expected for a
topological ``charge density''.) For finite-action fields $Q\in Z$. In terms
of the $\left(  \phi,\hat{n}\right)  $ parametrization (\ref{uparam}) for
$U\in SU\left(  2\right)  $, Eq. (\ref{q}) reduces to%
\begin{equation}
Q\left[  \phi,\hat{n}\right]  =\frac{1}{2^{4}\pi^{2}}\int d^{3}x\left(
\cos\phi-1\right)  \varepsilon_{ijk}\varepsilon^{abc}\partial_{i}\phi\hat
{n}^{a}\partial_{j}\hat{n}^{b}\partial_{k}\hat{n}^{c}.\label{qomn}%
\end{equation}

Two additional topological quantum numbers of $U$ are carried solely by its
$\hat{n}$-field component. The first is the homotopy degree of the maps
$\hat{n}^{a}$ from the space boundary $S_{\infty}^{2}$ (where $\left|  \vec
{x}\right|  \rightarrow\infty$) into the unit sphere $S_{\hat{n}}^{2}$ on
which $\hat{n}^{a}$ takes values. Continuous maps of this sort are classified
by the elements of the homotopy group $\pi_{2}\left(  S^{2}\right)  =Z$. (The
same topology characterizes the magnetic charge of Wu-Yang monopoles
\cite{wuy75}.) An explicit integral representation of their degree is
\begin{equation}
q_{m}\left[  \hat{n}\right]  =\frac{1}{8\pi}\int_{\partial R^{3}}d\sigma
_{i}\varepsilon_{ijk}\varepsilon^{abc}\hat{n}^{a}\partial_{j}\hat{n}%
^{b}\partial_{k}\hat{n}^{c}\label{qm}%
\end{equation}
where the integral extends over the closed surface $\partial R^{3}=S_{\infty
}^{2}$ at $\left|  \vec{x}\right|  \rightarrow\infty$. As expected, $q_{m}$ is
$\phi$-independent.

The third topological invariant owes its existence to the fact that all
$\hat{n}$ fields of finite action are required to approach a constant unit
vector at spacial infinity. As above, this requirement compactifies $R^{3}$
into $S_{sp}^{3}$ and thereby turns $\hat{n}\left(  \vec{x}\right)  $ into
continuous maps $S_{sp}^{3}\rightarrow S_{\hat{n}}^{2}$. Such maps carry a
Hopf charge $q_{H}$ which labels the elements of the homotopy group $\pi
_{3}\left(  S^{2}\right)  =Z$, i.e. the Hopf bundle \cite{ati90}. An explicit
integral representation for $q_{H}$ can be constructed by means of the local
isomorphism between the nonlinear $O\left(  3\right)  $ and $CP\left(
1\right)  $ fields which expresses $\hat{n}^{a}$ in terms of a complex
2-component field $\chi=\left(  \chi_{1},\chi_{2}\right)  $ with unit modulus
$\chi^{\dagger}\chi=\chi_{1}^{\ast}\chi_{1}+\chi_{2}^{\ast}\chi_{2}=1$ as
$\hat{n}^{a}=\chi^{\dagger}\tau^{a}\chi$. Then one has
\begin{equation}
q_{H}\left[  \hat{n}\left(  \chi\right)  \right]  =\frac{1}{4\pi^{2}}\int
d^{3}x\varepsilon_{ijk}\left(  \chi^{\dagger}\partial_{i}\chi\right)  \left(
\partial_{j}\chi^{\dagger}\partial_{k}\chi\right)  .\label{qh}%
\end{equation}
(A \emph{local} integral representation for $q_{H}$ directly in terms of the
$\hat{n}$-field does not exist.) In contrast to the topological charges $Q$
and $q_{m}$ which are of winding-number type, the Hopf invariant $q_{H}$ is a
linking number. The underlying topological structure enables and classifies
the link and knot solutions to be encountered in Sec. \ref{fn}.

Finally, we recall that the $U$ field topology - as summarized in the
conserved topological quantum numbers $Q$, $q_{m}$ and $q_{H}$ - characterizes
not only the saddle point solutions but a much larger class of continuous
field\ configurations with finite and in some cases even infinite (see below)\ action.

\subsection{Bogomol'nyi bound \label{bogbnd}}

The distribution of the solutions to Eqs. (\ref{omeq}) and (\ref{neq}) over a
denumerably \ infinite set of topological charge sectors allows for the
existence of\ more saddle points than could be handled in practical
applications of the expansion (\ref{zspa}). Hence additional criteria are
required to select the most relevant saddle points in a controlled fashion.
Such criteria will be established below, in the form of action bounds which
are monotonically increasing functions of the absolute topological charge
values. These bounds imply that contributions from saddle points with higher
topological quantum numbers to functional integrals are increasingly
suppressed by the Boltzmann factor $\exp\left(  -\Gamma\right)  $ and can be
systematically neglected. In the present section we establish the action bound
for fields which carry finite values of $Q$. A similar bound for fields with
nonvanishing Hopf charge will be obtained in Sec. \ref{fn}.

The lower bound on the action of the Lagrangian (\ref{l24d}) for any field $U$
with a well-defined homotopy degree (\ref{q}) can be derived from the
Lie-algebra valued expression%
\begin{equation}
M_{ij}=a\partial_{i}L_{j}+b\varepsilon_{ijk}L_{k}%
\end{equation}
where $a,b\in R$ and $L_{i}$ are the components of the Maurer-Cartan one-form
(\ref{L}). With the help of the Maurer-Cartan identity%
\begin{equation}
\partial_{i}L_{j}-\partial_{j}L_{i}=\left[  L_{j},L_{i}\right]
\end{equation}
one obtains for its square%
\begin{equation}
M_{ij}M_{ij}=a^{2}\partial_{i}L_{j}\partial_{i}L_{j}-2ab\varepsilon_{ijk}%
L_{i}L_{j}L_{k}+2b^{2}L_{i}L_{i}%
\end{equation}
which obeys the basic inequality
\begin{equation}
tr\left\{  M_{ij}M_{ij}\right\}  \leq0.\label{trmm}%
\end{equation}
(Recall from Eq. (\ref{L}) that the $L_{i}$ are expanded into anti-hermitean generators.)

After specializing the bound (\ref{trmm}) to $b=1/\sqrt{2}$ and $a=\pm
1/\left(  \sqrt{2}\mu\right)  $ \footnote{Different choices for the
coefficients $c_{0,1}$ in Eq. (\ref{gexp}) could be accomodated by modifying
the values of $a$ and $b$ and would result in a different factor in front of
$\left|  Q\right|  $ in the Bogomol'nyi bound.}, multiplying by $-\mu/\left(
2g^{2}\right)  $, integrating over $\vec{x}$ and using the integral
representation (\ref{q}) for $Q$, one arrives at%
\begin{equation}
-\frac{\mu}{2g^{2}\left(  \mu\right)  }\int d^{3}xtr\left\{  L_{i}%
L_{i}+\frac{1}{2\mu^{2}}\partial_{i}L_{j}\partial_{i}L_{j}\right\}  \geq
\mp\frac{12\pi^{2}}{g^{2}\left(  \mu\right)  }Q\left[  U\right]
.\label{prebnd}%
\end{equation}
The more stringent of these inequalities results from the lower (upper) sign
on the right-hand side if $Q>0$ ($Q<0$). Their left-hand side amounts to the
action which is produced by the first two terms \footnote{For special
purposes, e.g. for the discussion of infinite-action solutions, it may be
useful to maintain the surface term (\ref{sfterm}) in $\Gamma$. After
specializing to $b=1/\sqrt{2}$ and $a=\pm1/\left(  \sqrt{2}\mu\right)  $, the
obvious identity $tr\left\{  \partial_{i}L_{j}\partial_{i}L_{j}\right\}
=tr\left\{  \frac{1}{2}\partial^{2}\left(  L_{i}L_{i}\right)  -L_{i}%
\partial^{2}L_{i}\right\}  $ then results in $\Gamma\left[  U\right]
=-\frac{\mu}{2g^{2}\left(  \mu\right)  }\int d^{3}xtr\left\{  L_{i}%
L_{i}-\frac{1}{2\mu^{2}}L_{i}\partial^{2}L_{i}\right\}  \geq\mp\frac{12\pi
^{2}}{g^{2}\left(  \mu\right)  }Q\left[  U\right]  +\int d^{3}x\Delta
\mathcal{L}\left(  \vec{x}\right)  .$} of the Lagrangian (\ref{efflagr}).
Hence the expressions (\ref{prebnd}) for both signs combine into an inequality
of Bogomol'nyi type,%
\begin{equation}
\Gamma\left[  U\right]  \geq\frac{12\pi^{2}}{g^{2}\left(  \mu\right)  }\left|
Q\left[  U\right]  \right|  .\label{bb}%
\end{equation}
This is the desired lower bound on the action of any sufficiently smooth $U$
field with well-defined degree $Q$. It is saturated by those fields which obey
the Bogomol'nyi-type equation
\begin{equation}
\partial_{i}L_{j}=\mp\mu\varepsilon_{ijk}L_{k},\label{beq}%
\end{equation}
where the lower (upper) sign again refers to $Q>0$ ($Q<0$). The equations
(\ref{beq}) can be considered as analogs of the self-(anti)-duality equations
of Yang-Mills theory and have the interesting consequence that the rescaled
Maurer-Cartan currents $\tilde{L}_{i}:=\mp L_{i}/\left(  2i\mu\right)  $ of
minimal-action fields in any $Q$-sector become generators of the $su\left(
2\right)  $ Lie algebra,
\begin{equation}
\left[  \tilde{L}_{i},\tilde{L}_{j}\right]  =i\varepsilon_{ijk}\tilde{L}_{k}.
\end{equation}
Translationally invariant solutions (cf. Sec. \ref{classvac}), for example,
have $L_{i}=0$ and therefore trivially saturate the bound in the $Q=0$ sector.
It remains to be seen whether nontrivial solutions in sectors with larger
$\left|  Q\right|  $ exist as well \footnote{The analogous bound in the Skyrme
model for $Q=1$ can only be saturated on a hyperspherical domain $S^{3}$, for
example \cite{man86}. }

The large factor multiplying $\left|  Q\right|  $ in the inequality (\ref{bb})
indicates that contributions from saddle points with higher $\left|  Q\right|
$ are strongly suppressed. In fact, they seem safely negligible in most
amplitudes which receive nonvanishing contributions from the $Q=0$ sector.
However, one should keep in mind that even contributions with extremely small
``fugacities'' can sometimes have an important qualitative impact on the
partition function. A case in point are the decisive monopole contributions in
the 2+1 dimensional Yang-Mills-Higgs model \cite{pol77}. The physical
interpretation of the $\left|  Q\right|  \neq0$ solutions and their analogs in
Yang-Mills theory will be discussed in Sec. \ref{qspi}.

\section{Important saddle point solution classes}

\label{solncl}

In the following sections we are going to solve the four saddle point
equations (\ref{omeq}) and\ (\ref{neq}) explicitly. As already mentioned, our
main focus will be on solutions with a relatively high amount of symmetry
since their typically smaller action values\ enhance their contributions to
the saddle point expansion. Besides playing a predominant role in most
amplitudes, these solutions can often be obtained either analytically or with
moderate numerical effort.

\subsection{Translationally invariant solutions}

\label{classvac}

The simplest solutions of the saddle point equations (\ref{omeq}) and
(\ref{neq}) are the $\vec{x}$-independent matrices%
\begin{equation}
U_{c}=\exp\left[  \phi_{c}\hat{n}_{c}^{a}\frac{\tau^{a}}{2i}\right]  =const.
\end{equation}
where $\phi_{c}$ and $\hat{n}_{c}^{a}$ are both constant. These solutions form
the complete vacuum manifold of the dynamics (\ref{l24d}), i.e. the set of all
fields which attain the absolute action minimum%
\begin{equation}
\Gamma\left[  U_{c}\right]  =0.
\end{equation}
Due to a redundancy in the parametrization (\ref{uparam}), the subset of vacua
in the center of the gauge group is completely $\hat{n}^{a}$-independent:
\begin{equation}
\phi_{c,k}=2k\pi,\text{ \ \ \ \ }U_{c,k}=\left(  -1\right)  ^{k}.\label{omck}%
\end{equation}
In addition, those are the only vacua which do not break the global $U\left(
2\right)  $ symmetry of the Lagrangian (\ref{l24d}) spontaneously. A glance at
the integral representation (\ref{qomn}) shows that none of the $U_{c}$
carries topological charges $Q\neq0$.

\subsection{Constant-$\hat{n}$ solutions}

\label{constn}

Any constant vector $\hat{n}^{a}$ solves the saddle point equation (\ref{neq})
identically and reduces the other one, Eq. (\ref{omeq}) for $\phi$, to the
linear field equation%
\begin{equation}
\partial^{2}\left(  \partial^{2}\phi-2\mu^{2}\phi\right)
=0.\label{omeqconstn}%
\end{equation}
Obviously, the solutions of this equation constitute families of new saddle
points which differ by an additive constant and have degenerate action values.
(This trivially ensures periodicity in $\phi$.) Alternatively, Eq.
(\ref{omeqconstn}) and the action of its solutions can be derived from the
reduced Lagrangian \footnote{Although this Lagrangian is bilinear in $\phi$,
its contributions to $Z$ remain nontrivial since the Haar measure generates
interactions among the $\phi$ modes.}%
\begin{equation}
\mathcal{L}^{\left(  \hat{n}=c\right)  }=\frac{1}{2^{3}g^{2}\left(
\mu\right)  \mu}\left[  \left(  \partial^{2}\phi\right)  ^{2}+2\mu^{2}\left(
\partial_{i}\phi\right)  ^{2}\right]  .\label{lconstn}%
\end{equation}
The presence of the 4th-order term in Eq. (\ref{omeqconstn}) allows for
solution types which have no equivalent in Skyrme models. (Indeed, the
commutator or Skyrme term (cf. Eq. (\ref{commplus})) alone leads to a Laplace
equation for $\phi$, without a mass scale and with only constant regular
finite-action solutions on $R^{3}$.) A glance at the integrals \ (\ref{qomn})
- (\ref{qh}) shows that solutions with constant $\hat{n}$ do not carry any
topological quantum numbers, i.e.
\begin{equation}
Q^{\left(  \hat{n}=c\right)  }=q_{m}^{\left(  \hat{n}=c\right)  }%
=q_{H}^{\left(  \hat{n}=c\right)  }=0.
\end{equation}

All solutions of the linear field equation (\ref{omeqconstn}) can be
constructed by standard Green function techniques. The perhaps most
straightforward approach is to fold the static Klein-Gordon propagator%
\begin{equation}
\Delta\left(  \vec{x}-\vec{y};m\right)  =-\int\frac{d^{3}k}{\left(
2\pi\right)  ^{3}}\frac{e^{i\vec{k}\left(  \vec{x}-\vec{y}\right)  }}{\vec
{k}^{2}+m^{2}}=-\frac{1}{4\pi}\frac{e^{-m\left|  \vec{x}-\vec{y}\right|  }%
}{\left|  \vec{x}-\vec{y}\right|  }%
\end{equation}
with a ``scalar potential'' $\Phi$ which is defined both to be a solution of
the Laplace equation, $\partial^{2}\Phi=0$, and to act as the inhomogeneity of
the static Klein-Gordon equation
\begin{equation}
\partial^{2}\phi-2\mu^{2}\phi=\Phi.\label{inhomkg}%
\end{equation}
The field equation (\ref{omeqconstn}) is recovered from Eq. (\ref{inhomkg}) by
applying the Laplacian to both sides. Since the potential $\Phi$ plays the
role of a source for the $\phi$ field, inversion of the Klein-Gordon operator
immediately yields the general solution
\begin{equation}
\phi\left(  \vec{x}\right)  =\int d^{3}y\Delta\left(  \vec{x}-\vec{y};\sqrt
{2}\mu\right)  \Phi\left(  \vec{y}\right)  .\label{gensolconstn}%
\end{equation}
Of course, the regular finite-action solutions form but a small subset of
those comprised in Eq. (\ref{gensolconstn}).

Spherically symmetric solutions can be obtained more directly by restricting
the angular dependence of $\phi$, i.e. by substituting the ansatz $\phi\left(
r\right)  $ with $r:=\left|  \vec{x}\right|  $ into the general equation
(\ref{omeqconstn}) and ignoring for the moment potential singularities at the
origin. This yields the radial equation
\begin{equation}
r\phi^{\prime\prime\prime\prime}+4\phi^{\prime\prime\prime}-2\mu^{2}\left(
r\phi^{\prime\prime}+2\phi^{\prime}\right)  =0
\end{equation}
(radial derivatives $d/dr$ are denoted by a prime) whose four linearly
independent solutions can be found analytically:
\begin{equation}
\phi_{1}=c,\text{ \ \ \ \ }\phi_{2}=\frac{1}{\sqrt{2}\mu r},\text{
\ \ \ \ }\phi_{3}=\frac{e^{+\sqrt{2}\mu r}}{\sqrt{2}\mu r},\text{
\ \ \ \ }\phi_{4}=\frac{e^{-\sqrt{2}\mu r}}{\sqrt{2}\mu r}.
\end{equation}
The subset of regular finite-action solutions is therefore of the form%
\begin{equation}
\bar{\phi}^{\left(  \hat{n}=c\right)  }\left(  r\right)  =c_{1}+\frac{c_{2}%
}{\sqrt{2}\mu r}\left(  1-e^{-\sqrt{2}\mu r}\right)  .\label{phineqc}%
\end{equation}
The associated potential%
\begin{equation}
\bar{\Phi}^{\left(  \hat{n}=c\right)  }\left(  r\right)  =\left(  \partial
^{2}-2\mu^{2}\right)  \bar{\phi}^{\left(  \hat{n}=c\right)  }\left(  r\right)
=-2\mu^{2}\left(  c_{1}+\frac{c_{2}}{\sqrt{2}\mu r}\right)
\end{equation}
shows that the expression (\ref{phineqc}) in fact solves a generalization of
the homogeneous field equation (\ref{omeqconstn}), with an additional
delta-function singularity at the origin. Eq. (\ref{phineqc}) is therefore a
solution of Eq. (\ref{omeqconstn}) everywhere except at $\vec{x}=0$ and,
strictly speaking, one of its Green functions. A representative of this
solution class is drawn in Fig. 1.

After insertion into Eq. (\ref{gamsoft}), based on the Lagrangian
(\ref{l24d}), and use of the virial theorem (\ref{vth}) one finds the
solutions (\ref{phineqc}) to have the action
\begin{equation}
\Gamma\left[  \phi^{\left(  \hat{n}=c\right)  },\hat{n}_{c}\right]
=\frac{2\pi\mu}{g^{2}\left(  \mu\right)  }\int_{0}^{\infty}dr\left(
r\phi^{\left(  \hat{n}=c\right)  \prime}\right)  ^{2}=\frac{\pi}{\sqrt{2}%
}\frac{c_{2}^{2}}{g^{2}\left(  \mu\right)  }.
\end{equation}
This action is not subject to topological bounds and reaches the absolute
minimum $\Gamma=0$ for $c_{2}=0$ where the constant-$\hat{n}$ solutions turn
into the translationally invariant vacua of Sec. \ref{classvac}. (In contrast
to the center elements (\ref{omck}), however, the value of $\phi$ remains
unrestricted here.) Due to their partly very small action values, the
constant-$\hat{n}$ solutions may have a strong impact on the saddle point
expansion which should be explored in detail by studying their contributions
to suitable amplitudes.

\subsection{Faddeev-Niemi type knot solutions}

\label{fn}

In addition to the translationally invariant saddle points of Sec.
\ref{classvac}, there are other and less trivial solutions\ of the field
equations (\ref{omeq}) and\ (\ref{neq}) with constant $\phi$ fields. Among
them, the most intriguing class has the general form
\begin{equation}
\phi_{c2,k}=\left(  2k+1\right)  \pi,\text{ \ \ \ \ }U_{c2,k}\left(  \vec
{x}\right)  =\left(  -1\right)  ^{k}i\tau^{a}\hat{n}^{a}\left(  \vec
{x}\right)  ,
\end{equation}
which satisfies Eq. (\ref{omeq}) identically and carries no topological charge
$Q$. In fact, the nontrivial topology of $U_{c2}$ (as that of any other
constant-$\phi$ field configuration) has to reside exclusively in the $\vec
{x}$-dependence of its $\hat{n}$ field, whose dynamics is governed by the
($k$-independent) equation
\begin{equation}
\varepsilon^{abc}\partial_{i}\left[  \hat{n}^{b}\left(  2\mu^{2}\partial
_{i}\hat{n}^{c}+2\left(  \partial_{j}\hat{n}^{d}\right)  ^{2}\partial_{i}%
\hat{n}^{c}-4\partial_{j}\hat{n}^{c}\partial_{i}\hat{n}^{d}\partial_{j}\hat
{n}^{d}-\partial^{2}\partial_{i}\hat{n}^{c}\right)  \right]  =0.\label{knoteq}%
\end{equation}
The field equation (\ref{knoteq}) follows from the general saddle point
equation\ (\ref{neq}) by substituting $\phi_{c2.k}$ and simultaneously plays
the role of a (static) continuity equation for the conserved $O\left(
3\right)  $ current. Alternatively, it can be obtained by directly varying the
reduced Lagrangian%
\begin{equation}
\mathcal{L}^{\left(  \phi_{c2}\right)  }\left(  \vec{x}\right)  =\frac{\mu
}{g^{2}\left(  \mu\right)  }\left[  \left(  \partial_{i}\hat{n}^{a}\right)
^{2}+\frac{1}{\mu^{2}}\left(  \varepsilon^{abc}\partial_{i}\hat{n}^{b}%
\partial_{j}\hat{n}^{c}\right)  ^{2}+\frac{1}{2\mu^{2}}\left(  \varepsilon
^{abc}\hat{n}^{b}\partial^{2}\hat{n}^{c}\right)  ^{2}\right]  \label{ln}%
\end{equation}
which follows from Eq.\ (\ref{l24d}) after specialization to $\phi_{c2.k}$ and
reproduces the\ action (\ref{gamsoft}) for the $U_{c2,k}$. Remarkably, the
Lagrangian (\ref{ln}) is a generalization of the static Skyrme-Faddeev-Niemi
(SFN) model \cite{fad70,fad97}
\begin{equation}
\mathcal{L}^{\left(  SFN\right)  }=\frac{1}{2\lambda^{2}}\left(  \partial
_{i}\hat{n}^{a}\right)  ^{2}+\frac{e^{2}}{2}\left(  \varepsilon^{abc}%
\partial_{i}\hat{n}^{b}\partial_{j}\hat{n}^{c}\right)  ^{2}.\label{fnl}%
\end{equation}
In contrast to the SFN model, however, which was postulated on the basis of
qualitative symmetry and renormalization group arguments \cite{fad70,fad97},
our Lagrangian (\ref{ln}) follows uniquely from the Yang-Mills dynamics and
the Gaussian approximation to the vacuum wave functional. All coefficients are
therefore fixed in terms of the IR scale $\mu$ and the coupling $g\left(
\mu\right)  $, i.e. Eq. (\ref{ln}) does not contain free parameters.

Particular solutions of equation (\ref{knoteq}) are $\hat{n}^{a}=const.$,
which belong to the class of translationally invariant vacua (cf. Sec.
\ref{constn}), and $\hat{n}^{a}=\hat{x}^{a}$ (except at $r=0$) which is an
example from the ``hedgehog'' solution family whose detailed discussion will
be the subject of the following sections. The $\phi_{c2}$ hedgehog has
infinite action (since $U_{2c,k}$ develops a monopole-type singularity at
$\vec{x}=0,$ cf. Sec. \ref{shs}) and its Lagrangian reduces exactly to the
Faddeev-Niemi form (\ref{fnl}). Eq. (\ref{knoteq}) does probably also have
cylindrically symmetric vortex solutions which are analogs of the ``baby
Skyrmion'' solutions \cite{pie95} in similar models.

The most interesting and many-faceted solution classes of the field equation
(\ref{knoteq}), however, are expected to be twists, linked loops and knots
made of closed fluxtubes. Indeed, an intriguing variety of such topological
soliton solutions was found numerically\ for the SFN part (\ref{fnl}) of the
Lagrangian (\ref{ln}) in Refs. \cite{fad97,bat98}. These solutions generally
lack axial symmetry and carry a finite Hopf charge \footnote{Despite the
topological stability of the solitons with nonvanishing $q_{H}$ in the
sub-model (\ref{ln}), it might be possible to ``unwind'' and thus destabilize
them by excitations into the $\phi$ direction of the complete theory. This
possibility deserves further investigation.} $q_{H}^{\left(  c2\right)  }%
\neq0$. Moreover, their number and complexity increases strongly with the
value of $\left|  q_{H}^{\left(  c2\right)  }\right|  $. As in the
higher-$\left|  Q\right|  $ solution sectors discussed previously, a
practically useful saddle point expansion thus requires an effective means for
selecting the relevant contributions in a controlled fashion.

As anticipated in Sec. \ref{bogbnd}, such a means can be provided by
establishing that the action $\Gamma^{\left(  \phi_{c2}\right)  }$ based on
the Lagrangian (\ref{ln}) is bounded from below by a monotonically increasing
function of $\left|  q_{H}\right|  $. Actually, this just requires a
straightforward adaptation of a known bound on the SFN action \cite{vak79}.
One combines the obvious inequalities
\begin{align}
\Gamma\left[  \phi_{c2},\hat{n}\right]   &  \geq\frac{\mu}{g^{2}\left(
\mu\right)  }\int d^{3}x\left[  \left(  \partial_{i}\hat{n}^{a}\right)
^{2}+\frac{1}{\mu^{2}}\left(  \varepsilon^{abc}\partial_{i}\hat{n}^{b}%
\partial_{j}\hat{n}^{c}\right)  ^{2}\right]  \\
&  \geq\frac{2}{g^{2}\left(  \mu\right)  }\left[  \int d^{3}x\left(
\partial_{i}\hat{n}^{a}\right)  ^{2}\right]  ^{1/2}\left[  \int d^{3}x\left(
\varepsilon^{abc}\hat{n}^{a}\partial_{i}\hat{n}^{b}\partial_{j}\hat{n}%
^{c}\right)  ^{2}\right]  ^{1/2}%
\end{align}
(the first one holds because the omitted term in the Lagrangian (\ref{ln}) is
manifestly non-negative; the second one is a consequence of $\left(
a-b\right)  ^{2}\geq0$ for any real $a,b$) with the Sobolev-type inequality
\cite{vak79}%
\begin{equation}
\left|  q_{H}\left[  \hat{n}\right]  \right|  ^{3/2}\leq\frac{1}{2^{6}\sqrt
{2}3^{3/4}\pi^{4}}\int d^{3}x\sqrt{\left(  \varepsilon^{abc}\hat{n}%
^{a}\partial_{i}\hat{n}^{b}\partial_{j}\hat{n}^{c}\right)  ^{2}}\int
d^{3}x\left(  \varepsilon^{abc}\hat{n}^{a}\partial_{i}\hat{n}^{b}\partial
_{j}\hat{n}^{c}\right)  ^{2}\label{soi}%
\end{equation}
and a simple inequality due to Ward \cite{war99},
\begin{equation}
\left[  \left(  \partial_{i}\hat{n}^{a}\right)  ^{2}\right]  ^{2}\geq2\left(
\varepsilon^{abc}\hat{n}^{a}\partial_{i}\hat{n}^{b}\partial_{j}\hat{n}%
^{c}\right)  ^{2},
\end{equation}
to end up with the bound%
\begin{equation}
\Gamma\left[  \phi_{c2},\hat{n}\right]  \geq\frac{2^{9/2}3^{3/8}\pi^{2}}%
{g^{2}\left(  \mu\right)  }\left|  q_{H}\left[  \hat{n}\right]  \right|
^{3/4}.\label{vkb}%
\end{equation}
($2^{9/2}3^{3/8}\pi^{2}\simeq337.\,\allowbreak17$) This bound is rather rough
and could probably be made more stringent by incorporating the 3rd term of the
Lagrangian (\ref{ln}) and by refining the Sobolev bound (\ref{soi}) which is
expected to remain valid with about half of the factor on its right-hand side
\cite{war99}.

The probably most important lesson of the present section is a new physical
interpretation for Faddeev-Niemi-type knot solutions. In our framework, they
reemerge as gauge invariant IR degrees of freedom which represent gluon field
ensembles with a nonvanishing ``collective'' Hopf charge in the vacuum overlap
and other amplitudes. This new interpretation may actually put the tentative
identification of knot solutions with glueballs, advocated as a natural
generalization of the fluxtube picture for quark-antiquark mesons in Refs.
\cite{fad97,nie03}, on a more solid basis. Indeed, the original association of
Refs. \cite{fad97,nie03} is obscured by the interpretation of the $\hat{n}$
field as a gauge-dependent local color direction. Our $\hat{n}$ field, on the
other hand, is manifestly gauge invariant \footnote{An additional benefit of
the gauge invariant $\hat{n}$ field is that one does not have to deal with
unwanted, colored Goldstone bosons if the $O\left(  3\right)  $ symmetry of
the Lagrangians (\ref{l24d}) and (\ref{ln}) is spontaneously broken.}.
Moreover, glueballs are anyhow natural candidates for gluonic IR degrees of
freedom in the $Q=0$ sector, so that their (perhaps partial or indirect)
appearance in the saddle point solution spectrum would not be unexpected.

An additional advantage of our new framework for the knot dynamics is that it
allows the   investigation of potential relations between the solutions of Eq.
(\ref{knoteq}) and specific Yang-Mills fields which may play important roles
in the vacuum, including e.g. topologically nontrivial pure-gauge fields in a
non-linear maximally Abelian gauge \cite{van01} and center vortices
\cite{gre03}. If such relations exist, they could perhaps be qualitatively
traced by  analytical methods. A full quantitative survey of the knot solution
sector, however, will\ require a devoted numerical effort \footnote{Progress
towards analytical solutions of knot equations of the type (\ref{knoteq}) is
impeded by the fact that the obvious ans\"{a}tze tend to fail. A few numerical
knot solutions with small $\left|  q_{H}\right|  $ are probably sufficient for
the saddle point expansion, however, since contributions to physical
amplitudes are severely limited by the stringent action bound (\ref{vkb}),
especially when $Q=q_{H}=0$ solutions contribute as well.}.

\subsection{Hedgehog solutions}

\label{hhtype}

The existence of Skyrmions \cite{zah86} in nonlinear $\sigma$-models with
higher-derivative interactions suggests that our field equations (\ref{omeq})
and (\ref{neq}) have topological soliton solutions of ``hedgehog'' type,
\begin{equation}
\hat{n}^{a}\left(  \vec{x}\right)  =\hat{x}^{a},\text{ \ \ \ \ \ }\phi\left(
\vec{x}\right)  =\phi^{\left(  hh\right)  }\left(  r\right)  \label{hh}%
\end{equation}
($\hat{x}^{a}\equiv\vec{x}/r$, $r\equiv\left|  \vec{x}\right|  $), as well.
Their $SU\left(  2\right)  $ ``grand spin'' symmetry characterizes the
invariance of the corresponding $U$ fields under simultaneous spatial and
internal rotations and implies a substantial simplification of their dynamics.
(For larger gauge groups $SU\left(  N\right)  $ with $N>2$, the
three\ components of $\hat{x}$ form the part of the $\hat{n}^{a}$ field which
parametrizes the trivially embedded $SU\left(  2\right)  $ subgroup.) In the
present section, we discuss general properties of the hedgehog fields and
derive their reduced Lagrangian and saddle point equation. In the subsequent
sections \ref{anhh} and \ref{qspi}, we find the most important solution
classes explicitly and determine their physical interpretation.

The principal topological characteristic of the hedgehog configurations
(\ref{hh}) is their $\pi_{3}\left(  S^{3}\right)  $ winding number $Q$. For
fields of the form (\ref{hh}), its integral representation (\ref{qomn})
reduces to
\begin{align}
Q\left[  \phi^{\left(  hh\right)  }\right]    & =\frac{1}{2\pi}\int
_{0}^{\infty}\phi^{\left(  hh\right)  \prime}\left(  \cos\phi^{\left(
hh\right)  }-1\right)  dr\nonumber\\
& =\frac{1}{2\pi}\left[  \sin\phi^{\left(  hh\right)  }\left(  \infty\right)
-\sin\phi^{\left(  hh\right)  }\left(  0\right)  +\phi^{\left(  hh\right)
}\left(  0\right)  -\phi^{\left(  hh\right)  }\left(  \infty\right)  \right]
.\label{hq}%
\end{align}
(As expected from a topological invariant, it depends only on the boundary
values of the $\phi$ field.) In Sec. \ref{top} we established that finite
action fields carry integer values of $Q$, and Eq. (\ref{hq}) confirms this
explicitly. Indeed, the parametrization (\ref{uparam}) for hedgehog fields
(\ref{hh}) implies that well-defined $U$ fields necessitate the boundary
condition $\phi^{\left(  hh\right)  }\left(  0\right)  =2k_{1}\pi$ and that
finite-action fields must additionally satisfy $\phi^{\left(  hh\right)
}\left(  \infty\right)  =2k_{2}\pi$ (see below) where $k_{1,2}$ and thus
$Q=k_{1}-k_{2}$ are integers. Nevertheless, it is instructive to consider the
more general boundary conditions
\begin{equation}
\phi^{\left(  hh\right)  }\left(  0\right)  =n\pi,\text{ \ \ \ \ }%
\phi^{\left(  hh\right)  }\left(  \infty\right)  =m\pi,\text{ \ \ \ \ }%
Q\left[  \phi^{\left(  hh\right)  }\right]  =\frac{n-m}{2}\label{q-hh}%
\end{equation}
($n,m$ integer) which admit infinite-action fields with half-integer winding
numbers as well (for either $m$ or $n$ odd) \footnote{The half-integer winding
numbers cannot be associated with the degree of a map. The latter is defined
for maps between \emph{compact} manifolds only, which in our context implies
finite action.}. We will show in Sec. \ref{qspi} that hedgehog solutions to
the saddle point equations (\ref{omeq}), (\ref{neq}) under the boundary
conditions (\ref{q-hh}), both with finite and infinite action, can indeed be
found. The hedgehog solutions with $Q=\pm1$ will be of particular importance
since they probably dominate all $Q\neq0$ contributions to the saddle point
expansion. This follows from the bound (\ref{bb}) and from Skyrme-model type
arguments \cite{zah86} which suggest that the minimal-action solutions in the
$\left|  Q\right|  =1$ sectors are  hedgehogs.

In addition, all hedgehog fields carry one unit of a second topological
quantum number, the monopole-type charge $q_{m}\in\pi_{2}\left(  S^{2}\right)
$. This becomes explicit when evaluating the ($\phi$ independent) integral
representation (\ref{qm}) for $q_{m}$ with $\hat{n}^{a}=\hat{x}^{a}$:
\begin{equation}
q_{m}^{\left(  hh\right)  }:=q_{m}\left[  \hat{x}\right]  =\frac{1}{4\pi}\int
d\sigma_{i}\frac{\hat{x}_{i}}{r^{2}}=1.
\end{equation}
Obviously, $q_{m}^{\left(  hh\right)  }$ is independent of the boundary
conditions for $\phi$ and therefore of $Q$. The field with $q_{m}^{\left(
hh\right)  }=-1$ (the ``anti-monopole'') is obtained by replacing $\hat{x}%
_{i}$ by $-\hat{x}_{i}$, which corresponds to $U\rightarrow U^{\dagger}$ for
fixed $\phi$. Eq. (\ref{qh}) reveals, finally, that the Hopf charge of all
hedgehog configurations vanishes.

The dynamics of the hedgehog fields is governed by the soft-mode Lagrangian
(\ref{l24d}). Since $\hat{x}$ is an identical solution of the general field
equation (\ref{neq}) for $\hat{n}$, Eq. (\ref{l24d}) can be directly
specialized to $\hat{n}^{a}=\hat{x}^{a}$. Hence the integration over angles
becomes trivial and the hedgehog action turns into%
\begin{equation}
\Gamma\left[  \phi^{\left(  hh\right)  },\hat{x}\right]  =\int_{0}^{\infty
}dr\mathcal{L}^{\left(  hh\right)  }\left(  r\right)
\end{equation}
where $\mathcal{L}^{\left(  hh\right)  }$ is a $\phi^{\left(  hh\right)  }%
$-dependent radial Lagrangian. After substituting the ansatz (\ref{hh}) into
the full Lagrangian (\ref{l24d}), dropping total derivatives, suppressing the
superscript of $\phi^{\left(  hh\right)  }$ and again denoting radial
derivatives $d/dr$ by a prime, one arrives at the explicit expression
\begin{equation}
\mathcal{L}^{\left(  hh\right)  }\left(  r\right)  =\frac{\pi}{g^{2}\left(
\mu\right)  \mu}\left[  \frac{1}{2}\left(  r\phi^{\prime\prime}\right)
^{2}+\left(  3+\mu^{2}r^{2}\right)  \left(  \phi^{\prime}\right)  ^{2}%
+4\mu^{2}\left(  1-\cos\phi\right)  \right]  .\label{lrad}%
\end{equation}
All terms in $\mathcal{L}^{\left(  hh\right)  }$ are nonnegative. This has the
consequence that each of them must vanish  individually at any absolute action
minimum. The complete set of hedgehog vacua is therefore $\left(  \phi
_{c,k}^{\left(  hh\right)  },\hat{n}^{\left(  hh\right)  }\right)  =\left(
2k\pi,\hat{x}\right)  $ and forms a subset of the translationally invariant
center elements (\ref{omck}). As anticipated, any finite-action solution of
the form (\ref{hh}) has to approach one of these constant minima when
$r\rightarrow\infty$. The constant solutions $\phi_{c2,k}^{\left(  hh\right)
}=\left(  2k+1\right)  \pi$, on the other hand, are maxima of the action. A
representative of this type was already encountered in Sec. \ref{fn}.

The radial\ equation for $\phi^{\left(  hh\right)  }$ can be derived by
inserting the ansatz (\ref{hh}) into the general field equation (\ref{omeq})
or, more directly, by varying the radial Lagrangian (\ref{lrad}). Either way,
the result is%
\begin{equation}
r^{2}\phi^{\prime\prime\prime\prime}+4r\phi^{\prime\prime\prime}-2\left(
2+\mu^{2}r^{2}\right)  \phi^{\prime\prime}-4\mu^{2}r\phi^{\prime}+4\mu^{2}%
\sin\phi=0,\label{eqrad}%
\end{equation}
i.e. an ordinary nonlinear differential equation of fourth order and of
Fuchsian type. The associated boundary value problem can be solved numerically
with rather moderate computer resources. The exploration of the full solution
space is aided by two discrete symmetries of Eq. (\ref{eqrad}) which imply
that any solution $\bar{\phi}\left(  r\right)  $ gives rise to the additional
solutions $-\bar{\phi}\left(  r\right)  $\ and $\bar{\phi}\left(  r\right)
+2n\pi$. The former is a consequence of $\Gamma\left[  U\right]
=\Gamma\left[  U^{\dagger}\right]  $ while the latter simply reflects the
periodicity in the angular variable $\phi$.

Not surprisingly, the field equation (\ref{eqrad}) comprises the Gribov
equation \cite{gri78}. It consists of the terms proportional to $\mu^{2}$
which originate from the nonlinear-$\sigma$-model part\ of the Lagrangian
(\ref{lrad}) and dominate when $\mu$ becomes the largest scale and/or when the
higher derivatives become small \footnote{In the transition region between the
boundary values, however, the $\mu$-independent terms of Eq. (\ref{eqrad}) are
not negligible. This makes stable soliton solutions possible.}. The analogy
between the nonlinear potential term $\propto\sin\phi$ and a one-dimensional
pendulum in a gravitational field \footnote{This analogy becomes more explicit
after replacing $r$ with the logarithmic variable $t=\ln\left(  r/r_{0}%
\right)  $.}, often used to characterize the solution spectrum of the Gribov
equation, therefore applies to Eq. (\ref{eqrad}) as well. The stable
(unstable) equilibrium positions of the ``pendulum''\ are $\phi=\pi$ ($\phi
=0$), modulo a multiple of $2\pi$ which represents additional full turns. It
will be shown in Sec. \ref{qspi} that this analogy suffices to understand the
qualitative behavior of all numerical solutions.

\subsection{Analytical hedgehog solutions by series expansion \label{anhh}}

The $\left(  \phi,\hat{n}\right)  $ parametrization (\ref{uparam}) of the $U$
field implies that regular solutions $\phi^{\left(  hh\right)  }$ of the
radial hedgehog equation (\ref{eqrad}) approach a multiple of $2\pi$ at the
origin. Their small-$r$ behavior can therefore be determined analytically,
either by expanding the nonlinearity of Eq. (\ref{eqrad}) into powers of small
deviations $\delta\phi\left(  r\right)  $ from the constant action minima
$\phi_{c,k}^{\left(  hh\right)  }=2k\pi$ or by expanding the $r$ dependence of
the full solution into a Frobenius series. Similarly, finite-action solutions
can be obtained for $r\rightarrow\infty$ by asymptotically expanding around
the hedgehog vacua. Inside their regions of validity, these expansions provide
useful insights into the qualitative behavior of the hedgehog solutions as
well as quantitative checks on the numerical solutions to be found in Sec.
\ref{qspi}.

We start by deriving the solutions of the linearized hedgehog equation and the
corresponding power series expansion around the origin. Inserting the ansatz
\begin{equation}
\phi\left(  r\right)  =2k\pi+\delta\phi\left(  r\right)  +O\left(  \delta
\phi^{2}\right)
\end{equation}
into the radial saddle point equation (\ref{eqrad}) and retaining only terms
up to first order in $\delta\phi$, one arrives at the fourth-order linear
differential equation
\begin{equation}
r^{2}\delta\phi^{\prime\prime\prime\prime}+4r\delta\phi^{\prime\prime\prime
}-2\left(  2+\mu^{2}r^{2}\right)  \delta\phi^{\prime\prime}-4\mu^{2}%
r\delta\phi^{\prime}+4\mu^{2}\delta\phi=0\label{lineq}%
\end{equation}
which can be solved analytically by standard techniques. The general solution
is a superposition of four linearly independent base solutions $\delta\phi
_{i}$,
\begin{equation}
\delta\phi\left(  r\right)  =\sum_{i=1}^{4}\tilde{c}_{i}\delta\phi_{i}\left(
r\right)  ,\label{linsoln}%
\end{equation}
whose real, dimensionless coefficients $\tilde{c}_{i}$ remain undetermined and
have to be specified by imposing initial or boundary conditions. The base
solutions $\left\{  \delta\phi_{i}\right\}  $ are
\begin{equation}
\delta\phi_{1}=\mu r,\text{ \ \ \ \ }\delta\phi_{2}=\frac{1}{\mu^{2}r^{2}%
},\text{ \ \ \ \ }\delta\phi_{3}=\left(  1+\sqrt{2}\mu r\right)
\frac{e^{-\sqrt{2}\mu r}}{\mu^{2}r^{2}},\text{ \ \ \ \ }\delta\phi_{4}=\left(
1-\sqrt{2}\mu r\right)  \frac{e^{+\sqrt{2}\mu r}}{\mu^{2}r^{2}}.\label{pbas}%
\end{equation}
The requirements of regularity and uniqueness on the solutions at the origin
dictate two of the boundary conditions. The first one, $\delta\phi\left(
0\right)  =0$, implies $\phi\left(  0\right)  =2k\pi$ and thus ensures
uniqueness at $r=0$ while the second one, $\delta\phi^{\prime\prime}\left(
0\right)  =0$, is then imposed by the behavior of the base solutions
(\ref{pbas}). Accordingly, the small-$r$ behavior of the general regular
solution is restricted to
\begin{equation}
\phi\left(  r\right)  \overset{r\rightarrow0}{\longrightarrow}2n\pi+c_{1}\mu
r+c_{2}\mu^{3}r^{3}+O\left(  \mu^{4}r^{4}\right)
\end{equation}
where the constants $c_{1,2}$ are linear combinations of the $\tilde
{c}_{1,3,4}$ whose values can e.g. be  specified by providing initial data for
$\phi^{\prime}\left(  0\right)  $ and $\phi^{\prime\prime\prime}\left(
0\right)  $. All higher-order coefficients of the expansion are then fixed.

Alternatively, one can obtain the solutions of the full, nonlinear saddle
point equation (\ref{eqrad}) towards $r\rightarrow0$ by analytical
continuation into a Frobenius series. A somewhat tedious calculation yields
\begin{equation}
\phi\left(  r\right)  =2n\pi+\phi_{1}r+\phi_{3}r^{3}+\frac{\mu^{2}}{14}\left(
\phi_{3}+\frac{\phi_{1}^{3}}{30}\right)  r^{5}+\frac{\mu^{2}}{2\cdot3^{3}%
7}\left[  \mu^{2}\left(  \phi_{3}+\frac{\phi_{1}^{3}}{30}\right)  +\frac{1}%
{2}\phi_{1}^{2}\phi_{3}-\frac{1}{5!}\phi_{1}^{5}\right]  r^{7}+O\left(
r^{9}\right)  \label{sersoln}%
\end{equation}
where the coefficients $\phi_{1}$ and $\phi_{3}$ are again left to be
determined by initial conditions. Even-order derivatives of $\phi$ (or
equivalently the coefficients $\phi_{2k}$) vanish at $r=0$ while those of odd
order, $\phi_{2k+1}$, can be expressed in terms of $\phi_{1}$ and $\phi_{3}$.
A comparison between Eqs. (\ref{linsoln}) and (\ref{sersoln}) shows that the
solutions of the exact radial equation start to differ from those of the
linearized equation at $O\left(  r^{3}\right)  $. Hence the series solution
(\ref{sersoln}) permits a more accurate check of the numerical solutions over
a larger radial interval.

Analogous expansions\ around the constant action minima $\phi_{c,k}^{\left(
hh\right)  }=2k\pi$ exist asymptotically, i.e. towards $r\rightarrow\infty$,
for all finite-action solutions. Infinite-action solutions of Eq.
(\ref{eqrad}), finally, can be linearized around the constant solutions
$\phi_{2,k}^{\left(  hh\right)  }=\left(  2k+1\right)  \pi$ which\ they
approach at\ one or both ends of the radial domain. The resulting equation for
$\delta\phi$ differs from Eq. (\ref{lineq}) only in the sign of the
$\delta\phi$ term. Its solutions are linear combinations of generalized
hypergeometric functions.

\section{Numerical analysis and physical interpretation}

\label{qspi}

We now turn to the numerical solution of the hedgehog saddle point equation
(\ref{eqrad}). Due to the periodicity in $\phi$, the considered range of
boundary values can be limited without loss of generality to $\phi\left(
0\right)  \in\left]  0,2\pi\right]  $. Regularity at the origin then further
specifies $\phi\left(  0\right)  =2\pi$ and imposes $\phi^{\prime\prime
}\left(  0\right)  =0$ (see Sec. \ref{anhh}). The value of the topological
charge $Q$ fixes a third boundary condition,
\begin{equation}
\phi\left(  \infty\right)  =2\pi\left(  1-Q\right)  ,\label{phihhinf}%
\end{equation}
owing to Eq. (\ref{hq}). Hence all regular hedgehog solutions in a given
$Q$-sector can be found by just varying the value of a fourth boundary
condition. In the following, we use the initial slope $\beta:=\phi^{\prime
}\left(  0\right)  $ for this purpose. At the end of the section, we will also
find irregular solutions with $\phi\left(  0\right)  =\pi$. 

\subsection{Instanton classes}

We begin our exploration of the hedgehog solution space by searching for the
regular finite-action solutions of Eq. (\ref{eqrad}) which, as established in
Sec. \ref{top}, carry integer values of $Q$. The numerical analysis shows (and
the pendulum analogy in Sec. \ref{shs} will explain) that only one solution of
this type exists in each $Q$ sector. In the simplest case, $Q=0$, this is just
the translationally invariant vacuum solution $\phi^{\left(  hh\right)  }%
=2\pi$.

For $Q=1$ we find the prototypical nontrivial hedgehog solution, depicted in
Fig. 2. In order to clarify its physical interpretation, we note that it
shares the $\pi_{3}\left(  S^{3}\right)  $ homotopy classification, encoded in
the topological charge $Q\left[  U\right]  $ of the relative gauge orientation
$U=U_{-}^{-1}U_{+}$, with the Yang-Mills instanton \cite{bel75}. Of course,
both also share the saddle point property, as the instanton minimizes the
classical Euclidean Yang-Mills action in the $Q=1$ sector. In order to trace
their association further, we inspect the relative gauge orientation
$U_{I,YM}$ of a Yang-Mills  instanton with size $\rho$. It is of hedgehog form
as well, and its $\vec{x}$-dependence in (Euclidean) temporal gauge is known
to be \cite{bit78}
\begin{equation}
\phi_{I,YM}\left(  r\right)  =-\frac{2\pi r}{\sqrt{r^{2}+\rho^{2}}%
}\label{omin}%
\end{equation}
in the parametrization (\ref{hh}). For a direct comparison with our $Q=1$
solution, we have included $\phi_{I,YM}$ with $\rho=2\mu^{-1}$ as the dashed
curve in Fig. 2 (and adapted it to our periodicity interval convention by
adding $2\pi$). The radial dependence of both can be seen to be surprisingly
similar. This implies that the dominant contributions from all $Q=1$ gauge
field orbits to the vacuum overlap have a relative gauge orientation close to
that of the Yang-Mills instanton and  indicates that the $Q=1$ hedgehog
solution primarily summarizes contributions from the instanton orbit. (Of
course, one would not expect exact agreement since our solutions contain
scale-symmetry breaking quantum corrections \footnote{In this context, it is
interesting to recall that the analogous scale symmetry breaking due to 1-loop
corrections around Yang-Mills instantons is unable to stabilize their size
distribution.} and contributions from other $Q=1$ gauge fields as well.)
Accordingly, and generalizing the above findings to multi-instanton solutions,
we will refer to the unique regular finite-action solution of Eq.
(\ref{eqrad}) with integer $Q$ as the ``$Q$-instanton class''.

Our hedgehog saddle point equation (\ref{eqrad}) and its instanton class
solutions derive  from the gradient-expanded soft-mode Lagrangian
(\ref{efflagr}). It is instructive to compare this approach with a variational
estimate of one-instanton contributions to the bare action (\ref{gammab}) in
Ref. \cite{bro299}. By approximately minimizing the bare action with
one-parameter families of trial functions similar to the instanton profile
(\ref{omin}) and using qualitative scaling properties, it was argued in Ref.
\cite{bro299} that radiative corrections can stabilize the instanton size. Our
exact saddle point solutions make the dynamical size stabilization manifest.
We have already traced the underlying mechanism to the virial theorem
(\ref{vth}) which is independent of most specific features of the soft-mode
dynamics and thus overcomes the chronic infrared instability of dilute
instanton gases \cite{cal78} in a rather generic way. For $\mu\simeq1.5$ GeV,
the size $\rho\simeq2\mu^{-1}$ \footnote{The dynamical mass scale $\mu$ and
the average instanton size might be related to an effective IR\ fixed point of
the type considered in Ref. \cite{shu95}.} of the 1-instanton class solution
agrees inside errors with the results of instanton liquid model \cite{shu295}
and lattice \cite{mic95} simulations. It also assures that the two leading
terms of the gradient expansion (\ref{gexp}) yield a sufficiently accurate
approximation to the instanton action (cf. the comments below Eq.
(\ref{sfterm})).

Our 1-instanton class profile function $\phi_{I}\left(  r\right)  $ is rather
similar to the one found by approximately minimizing the bare action
(\ref{gammab}) variationally \cite{bro299}. This indicates that the bulk of
the instanton's physics and size distribution is generated by soft modes, as
one would intuitively expect. Our approach therefore provides a well-adapted
and efficient framework for the treatment of these and other vacuum fields. In
contrast to variational approaches, furthermore, it allows to systematically
find \emph{all} saddle points exactly (including those which are not of
hedgehog form). Already in the hedgehog sector, for example, we will find
solutions with more complex and unprecedented shapes than Eq. (\ref{omin}).
Since there is little guidance for the choice of suitable trial functions in
these and other cases, such solutions would be difficult to find variationally.

According to Eq. (\ref{q-hh}), all monotonic hedgehog solutions with $Q>0$
($Q<0$) have negative (positive) slopes $\beta=\phi^{\prime}\left(  0\right)
$ at the origin. The anti-instanton class\ with $Q=-1$, in particular, results
from changing the sign of the instanton boundary value, $\beta_{\bar{I}}%
=-\beta_{I}$, and can be obtained without further calculation: it simply
results from the combined action of the two symmetry transformations
$\phi\rightarrow-\phi$ and $\phi\rightarrow\phi+4\pi$ on the instanton class
solution. Hence the $Q=\pm1$ instanton classes have degenerate action values,
precisely as their Yang-Mills counterparts.

Multi-instanton class solutions are characterized by an integer topological
charge $Q\geq2$. The modulus $\left|  \beta_{Q,I}\right|  $ of their
(negative) initial slope grows monotonically with $Q$, i.e.
\begin{equation}
\left|  \beta_{I,Q^{\prime}}\right|  >\left|  \beta_{I,Q}\right|  \text{
\ \ \ \ for \ \ \ }Q^{\prime}>Q.
\end{equation}
As in the 1-instanton case, the action-degenerate multi-antiinstanton classes
with $Q\leq-2$ can be constructed by flipping the sign of the corresponding
multi-instanton classes and adding $4\pi$. The initial slopes of the
multi-(anti)instanton solutions are therefore related by
\begin{equation}
\beta_{\bar{I},-Q}=-\beta_{I,Q}.
\end{equation}
The $\left|  Q\right|  \geq2$ hedgehog solutions correspond to special
arrangements of the underlying Yang-Mills multi-(anti)instantons. In fact, the
relative gauge orientation $U=U_{-}^{-1}U_{+}$ of a  multi-instanton
configuration is of hedgehog type only if all individual (anti)instantons are
centered at the origin. This raises the question whether $\left|  Q\right|
\geq2$ Yang-Mills instanton solutions with separated individual positions are
at least approximately represented by other solutions of the saddle point
equations (\ref{omeq}) and (\ref{neq}). Experience from Skyrme-type models,
whose analogous $\left|  Q\right|  \geq2$ Skyrmion solutions are well
approximated by rational \cite{lon05} or harmonic \cite{ioa99} maps, might
suggest that similar types of non-hedgehog field configurations approximate
higher-$Q$ solutions of Eqs. (\ref{omeq}) and (\ref{neq}) as well.

The action of the 1-instanton class solution is large ($\Gamma_{I}%
\sim220/g^{2}$ at$\ \mu=1$ GeV), in analogy with the large action of typical
Yang-Mills instantons, and its direct impact on the saddle point expansion is
therefore small \footnote{Note that the 1-instanton class does not saturate
the action bound (\ref{bb}) and hence does not solve the Bogomol'nyi-type
equation (\ref{beq}). This is in constrast to the Yang-Mills instanton which
is the absolute minimum of the Euclidean Yang-Mills action in the $Q=1$ sector
and therefore self-dual. }. Moreover, the action bound (\ref{bb}) implies that
higher-$Q$ instanton classes should be irrelevant for most amplitudes, with
potentially important  exceptions as mentioned in Sec. \ref{bogbnd}. Instanton
``liquid'' vacuum models (ILMs) \cite{sch98} are built on the same premise and
suggest that physically far more relevant contributions originate instead from
ensembles of instantons and anti-instantons with equal average densities. It
would be important to determine whether contributions of this sort are
approximately represented by nontrivial $Q=0$ solutions of Eqs. (\ref{omeq})
and (\ref{neq}) as well. In any case, our above results imply that they cannot
be of hedgehog type.

\subsection{Meron classes}

We now extend our search to hedgehog solutions with infinite action. Although
their relevance for the saddle point expansion is not obvious, one might
speculate that their infinite-action suppression could be overcome by some
additional mechanism (see below). Our chief motivation for discussing them
here, however, derives from their association with the infinite-action meron
solutions \cite{dea76} of the classical (Euclidean)\ Yang-Mills equation.

Hedgehog solutions with infinite action are far more the rule than the
exception. In fact, all regular ($\phi\left(  0\right)  =2\pi$) solutions of
Eq. (\ref{eqrad}) with initial slopes $\beta$ inbetween the discrete set of
instanton-class values $\beta_{I,Q}$ have infinite action since they approach
one of the constant fields $\phi_{M}\left(  \infty\right)  =\left(
2k+1\right)  \pi$ towards spacial infinity. The latter carry a nonzero action
density (cf. Eq. (\ref{lrad})) and furthermore imply that the corresponding,
asymptotic $U$ fields $U_{M}\left(  \left|  \vec{x}\right|  \rightarrow
\infty\right)  $ remain angle-dependent. Exactly the same behavior
characterizes the relative gauge orientations $U=U_{-}^{-1}U_{+}$ of
Yang-Mills merons in temporal gauge, which are of hedgehog form as well.
Furthermore, Eq. (\ref{phihhinf}) shows that solutions with $\phi\left(
\infty\right)  =\left(  2k+1\right)  \pi$ carry half-integer topological
charge $Q$, again as the Yang-Mills merons. In analogy with the
instanton-class solutions of the previous section, we will therefore call
these solutions ``$2Q$-meron classes''.

The profile function $\phi_{M}\left(  r\right)  $ of a typical 1-meron class
solution with $Q=1/2$ is drawn in Fig. 3. A direct comparison with the
corresponding profile of the Yang-Mills meron in temporal gauge,
\begin{equation}
\phi_{M,YM}\left(  r\right)  =\frac{1}{2}\lim_{\rho\rightarrow0}\phi
_{I,YM}\left(  r\right)  =-\pi\theta\left(  r\right)  \label{ommer}%
\end{equation}
(where $\theta$ is the step function), is complicated by the fact that the
classical meron is pointlike while our solutions incorporate quantum effects
which break dilatation symmetry and stabilize their size at a finite value.
Such effects are expected to smoothen the singularity of the Yang-Mills meron,
too, and probably cause our solution $\phi_{M}\left(  r\right)  $ to become
non-monotonic by overshooting in the transition region. We therefore draw the
Yang-Mills meron profile in Fig. 3 (dashed curve) with the finite size
$\rho=2\mu$ of the instanton class solutions (and adapt it to our periodicity
interval convention).

Our nomenclature for multi-meron classes with $\left|  Q\right|  >1/2$
includes only solutions with half-integer topological charge because solutions
with even ``meron number'' and correspondingly integer $Q$ coincide with the
$Q$-instanton classes. This is expected since the relative gauge orientation
$U_{M,YM}$ of the underlying Yang-Mills multi-merons is of hedgehog type only
if all individual merons sit on top of each other. Such configurations, when
carrying integer overall values of $Q$, coalesce into the corresponding
Yang-Mills instantons and those are represented by the $Q$-instanton class
solutions in our framework.

The behavior of the multi-meron class solutions with $Q\geq3/2$ is
qualitatively rather similar to that of the 1-meron class, although size and
strength scales may differ substantially. As an example, Fig. 4 shows a
typical 3-meron class solution. An important general property of all
solutions with half-integer $Q$ is that they come in families of continuously
varying sizes. As already alluded to, this is because their size depends on a
second mass scale $\beta_{M}\equiv\phi_{M}^{\prime}\left(  0\right)  $ (in
addition to $\mu$) and because solutions for all values of $\beta_{M}$ in the
finite intervals%
\begin{equation}
\beta_{I,Q}>\beta_{M,Q+1/2}>\beta_{I,Q+1}%
\end{equation}
(where $Q\geq0$, i.e. $\beta\leq0$, and $\beta_{I,Q=0}=0$ are implied) can be
found. As in the instanton sector, multi-anti-meron classes with negative $Q$
are obtained from the positive-$Q$ solutions by changing their sign and adding
$4\pi$.

The variable mass scale $\beta_{M}$ in the meron sectors implies that there
are infinitely more meron than instanton class solutions. This makes it
tempting to speculate that the meron ``entropy'' contributions to the weight
function of functional integrals over $U$ might be able to overcome the
infinite-action suppression. If so, it would shed new light on the physical
interpretation not only\ of our solutions but also of the Yang-Mills merons
themselves, whose potential role remains controversial. Furthermore, it would
suggest a modified saddle point expansion in which action and entropy are
minimized jointly. These issues deserve further investigation.

\subsection{Singular hedgehog solutions\label{shs}}

Above we have classified all regular hedgehog solutions, i.e. those which
satisfy the initial condition $\phi\left(  0\right)  =2\pi$. We are now going
to examine the remaining solution classes of the radial field equation
(\ref{eqrad}). Its members share the alternative initial condition
$\phi\left(  0\right)  =\pi$, may display a rather complex spacial structure
and are characterized by a monopole-type singularity at the origin, i.e. they
solve Eq. (\ref{eqrad}) everywhere except at $\vec{x}=0$.

In order to understand the qualitative behavior of both regular and irregular
hedgehog solution classes from a common perspective, it is useful to elaborate
on the analogy between the hedgehog equation (\ref{eqrad}) and the pendulum
equation which was mentioned in Sec. \ref{hhtype}. According to this analogy,
the instanton classes correspond to exactly $Q$ full turns of the pendulum,
where the sign of $Q$ indicates the direction of the rotation. The pendulum
mass starts in the unstable equilibrium position at time $t=\ln r=-\infty$
with just enough initial speed $\beta=\phi^{\prime}\left(  0\right)  $ to
finally end up there again for $t=\ln r=+\infty$. This analogy implies, in
particular, that there is exactly one regular hedgehog solution for each
integer $Q$ and that the constant $\phi=2\pi$ is the only regular solution
with $Q=0$. The meron class solutions start from the unstable equilibrium
position as well. However, their initial velocity $\beta$ is insufficient for
completing all turns in full. The last turn remains uncompleted, i.e. the
pendulum swings back, oscillates around and finally settles into the stable
equilibrium position. Hence all meron class solutions have half-integer $Q$
and are non-monotonic.

For the irregular solutions, on the other hand, the pendulum starts at the
stable equilibrium position $\phi=\pi$. When not provided with sufficient
initial speed $\beta$ to complete a full turn, it just performs damped
oscillations around $\phi=\pi$. The corresponding solution has $Q=0$ and is
depicted in Fig. 5. When $\left|  \beta\right|  $ is sufficiently large,
however, the pendulum can perform $Q$ full turns before settling into the
stable equilibrium position. As a consequence, all these solutions have
integer $Q\neq0$ and infinite action (cf. Eq. (\ref{lrad})). Pursuing the
analogy further, one would also expect irregular solutions which have an
initial value $\phi^{\prime}\left(  0\right)  $ exactly as needed to end up at
the unstable equilibrium position when $r\rightarrow\infty$ (after possibly
completing a number of full turns). Such configurations would carry a
half-integer topologically charge $Q$ and a finite action. For obvious reasons
they turn out to be highly sensitive to variations of the initial condition,
however, and therefore difficult to establish numerically.

In contrast to the instanton and meron classes, the singular hedgehog
solutions do not seem to have obvious analogs among the solutions of the
classical Yang-Mills equation. As opposed to the instanton (meron) classes,
furthermore, the irregular integer-$Q$ (half-integer-$Q$) solutions have
infinite (finite) action. This allows for nontrivial hedgehog solutions with
$Q=0$, which are necessarily irregular at the origin. The physical
interpretation of all irregular hedgehog solutions and their relevance for the
saddle point expansion remain to be clarified.

\section{Summary and conclusions}

\label{suc}

The main results of this paper are a practicable saddle point expansion for
the Yang-Mills vacuum overlap amplitude in terms of gauge invariant, local
matrix fields and the identification of new gluonic IR degrees of freedom in
this framework. After adopting a gauge-projected Gaussian approximation to the
vacuum wave functional, the IR degrees of freedom can be obtained explicitly
as the saddle points of a soft-mode action which gather contributions from
dominant gluon field families to soft Yang-Mills amplitudes and thus represent
collective properties of the Yang-Mills dynamics. Since their gauge invariant
definition makes no reference to specific amplitudes, furthermore, the IR
degrees of freedom are universal. They provide both new structural insights
into the organization of the low-energy Yang-Mills dynamics and the principal
input for a systematic saddle point expansion of soft amplitudes.

Our survey of the saddle point solution space uncovered a diverse spectrum of
IR degrees of freedom which carry several topological charges with associated
Bogomol'nyi-type action bounds and obey a virial theorem which guarantees
their scale stability. Solutions with a relatively high degree of symmetry
were obtained either analytically or with modest numerical effort. Since
solutions of this type are generally characterized by small action values and
hence play a dominant role in the saddle point expansion, we have investigated
their properties in some detail. Besides translationally invariant vacua and
analytical solutions with a fixed relative gauge orientation, we have found
topological solitons of hedgehog, (vortex) link and knot types.

Some of the IR degrees of freedom have a transparent physical interpretation
directly in terms of the underlying gluon fields. The contributions from the
gauge orbits of Yang-Mills instanton and merons, in particular, are gathered
by saddle point fields of hedgehog type which share their (integer or
half-integer) topological charge and represent vacuum tunneling processes in
the Hamiltonian formulation of non-Abelian gauge theory. Although our saddle
point solutions contain quantum effects and potentially relevant contributions
from other gauge fields, those in the instanton class turn out to reproduce
the relative gauge orientation between the in- and out-vacua of the Yang-Mills
instanton rather closely. The finite extent of our meron solution classes, on
the other hand, is generated by quantum effects which smoothen the
singularities of the classical, pointlike Yang-Mills merons. Nevertheless, our
meron classes turn out to share the infinite action of their Yang-Mills counterparts.

Among all those IR fields which carry one unit of topological (instanton)
charge, the single (anti-) instanton classes are expected to attain the
minimal action value. As a consequence of the action bound, they should
therefore dominate the saddle point expansion in all topological charge
sectors. Similar configurations emerged in a variational treatment along with
qualitative arguments in favor of their size stabilization. In our approach,
this stabilization is manifest in the exact instanton class solutions
themselves. In fact, their size turns out to be fixed at about twice the
inverse IR gluon mass scale and agrees inside errors with instanton liquid
model and lattice results. The underlying virial theorem and the soft gluon
mass generation therefore provide new insight into the mechanism by which the
chronic infrared diseases of dilute Yang-Mills instanton gases are overcome.

The sizes of the meron class solutions with half-integer topological charge
turn out to be of a more complex origin. Besides the dynamical gluon mass they
depend on a second, variable mass scale which is encoded in a boundary
condition. Hence meron classes exist within large and continuous size ranges
and consequently form a far more extensive solution family than the instanton
classes. This opens up the hypothetical\ possibility for their entropy to
overcome their  infinite action suppression in functional integrals. Such a
mechanism would not only help to clarify the physical impact of our solutions
but also shed new light on the still controversial role of the Yang-Mills
merons themselves. In addition to the instanton and meron classes, finally,
there exists a third class of hedgehog solutions which contains a
monopole-type singularity at the origin. These irregular solutions can carry
half-integer and integer (including zero) topological charges as well, and
generally have infinite action. Hence their potential physical relevance seems
to depend on the existence\ of additional mechanisms which could both smoothen
their singularity and overcome their infinite-action suppression.

Several other remarkable families of IR degrees of freedom turn out to be
represented by topological solitons of Faddeev-Niemi type, i.e. by
(potentially twisted) links and knots. In fact, our saddle point dynamics
\ contains a specific generalization of the Faddeev-Niemi Lagrangian and shows
explicitly how it is embedded in the Gaussian approximation to the Yang-Mills
vacuum wave functional. This puts Faddeev-Niemi theory into a new perspective,
as the effective\ dynamics of dominant sets of gauge field orbits with a
collective Hopf charge, and provides the underlying unit-vector field with a
manifestly gauge invariant meaning. The latter would well accord with the
tentative interpretation of knot\ solutions as glueballs by Faddeev, Niemi and
coworkers. This and other interpretations could be tested in our framework by
directly evaluating the impact of the knot saddle points on suitable
amplitudes, e.g. on glueball correlation functions.

More generally, the saddle point expansion allows the systematic calculation
of contributions from all relevant IR degrees of freedom to functional
integrals which represent soft Yang-Mills amplitudes. First calculations of
this type, focusing on fundamental vacuum properties including gluon
condensates and the topological susceptibility, are underway. In addition, our
approach makes it possible to analyze the gauge field content of any IR
variable individually by applying standard functional techniques to the
integrals over gluon fields with which they are associated. Investigations of
this sort would not only provide further structural insight into specific IR
degrees of freedom and their physical role but may also shed new light on the
dynamical mechanisms by which soft gauge fields organize themselves into
collective degrees of freedom.

The diverse topological properties of the IR variables demonstrate that the
gauge-projected Gaussian wave functional not only captures the
homotopy\ structure of the gauge group but also implements ``derivative''
topologies which further characterize the saddle point families. Due to the
typical robustness of such topological properties, the related results are
expected to remain at least qualitatively valid beyond the Gaussian
approximation. Moreover, the saddle point expansion engenders the means to
test this expectation quantitatively, by mapping out limitations of the
underlying vacuum wave functional in comparison with lattice data. Extensions
of our framework to suitable supersymmetric gauge theories would even permit
analytical tests of this sort, e.g. by tracing vestiges of the monopole-based
confinement mechanism in the vacuum functional. The insights gained from such
investigations may also  provide specific guidance for the development of
improved collective-mode actions and  consequently generate systematic
corrections to the IR variables.

Our IR saddle point expansion can be extended in several directions. A first
important task would be a more exhaustive survey of the saddle-point solution
space which should encompass potentially relevant approximate solutions. The
extension to QCD proper requires the generalization to the gauge group
$SU\left(  3\right)  $ and the implementation of quarks into the Gaussian wave
functional, both of which pose no conceptual problems. Most topological
properties, in particular, reside in the trivially embedded $SU\left(
2\right)  $ subgroup of the full color group and will remain unchanged. A
sufficiently complete treatment of the quark-gluon interactions and their
impact on the effective action, however, appears to be more challenging.

Our approach opens up a variety of directions for future research. Besides
those already mentioned, it would for example be interesting to explore
relations between the gauge-invariant IR degrees of freedom and
gauge-dependent gluonic structures (monopoles, vortices, branes etc.) and
amplitudes (e.g. the 2-dimensional nonlocal gluon condensate and Green
functions)\ which appear in gauge-fixed formulations. Another useful endeavor
would be the calculation of those higher-dimensional vacuum condensates which
provide the principal input for the operator product expansion of glueball
correlators. Duality sum rules could then link the contributions from
different IR degrees of freedom with the low-lying glueball spectrum
\cite{for05}, e.g. as a precursor and complement to a direct saddle-point
evaluation of glueball correlation functions.

\begin{acknowledgments}
\bigskip
\end{acknowledgments}

Financial support by the Conselho Nacional de Desenvolvimento Cient\'{\i}fico
e Tecnol\'{o}gico (CNPq) of Brazil as well as the hospitality and a visiting
scientist grant of the ECT* in Trento (Italy) are gratefully acknowledged.

\newpage

{\Large Figure captions:}

\begin{enumerate}
\item[Fig. 1: ] An example of the $\hat{n}=const.$ solution class. As in all
following figures, we display the solution at $\mu=1$ GeV. Other values of
$\mu$ can immediately be accommodated by scaling the $r$-axis. Our conventions
for the periodicity interval of $\phi$ restrict its initial value to
$\phi\left(  0\right)  \in\left]  0,2\pi\right]  $.

\item[Fig. 2: ] The canonical 1-instanton class solution with $Q=1$.

\item[Fig. 3: ] A typical meron-class solution with $Q=1/2$.

\item[Fig. 4: ] A 3-meron class solution with $Q=3/2$.

\item[Fig. 5: ] An example for a nontrivial $Q=0$ hedgehog solution with
monopole-type behavior at the origin.
\end{enumerate}

\bigskip

\bigskip
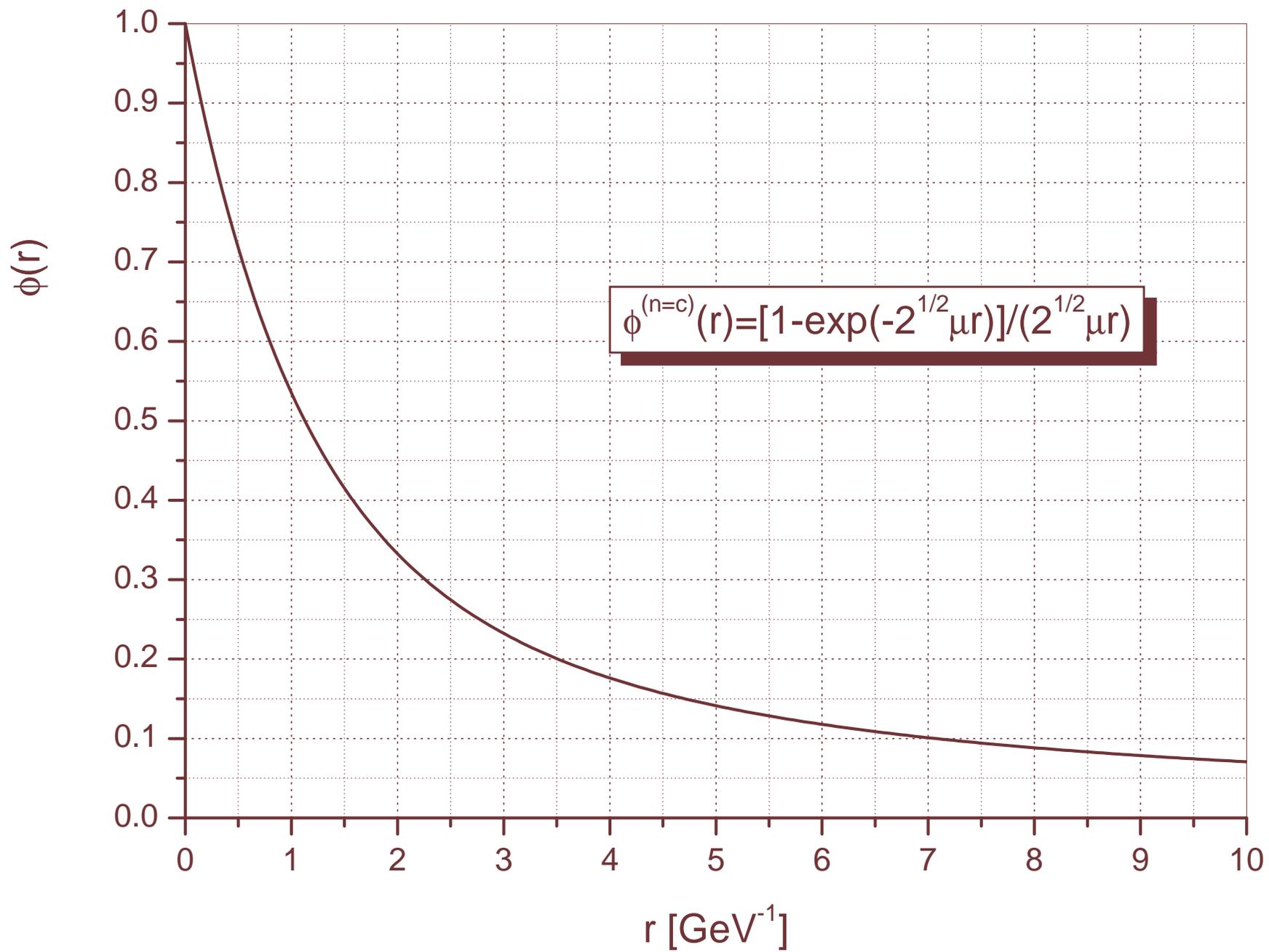
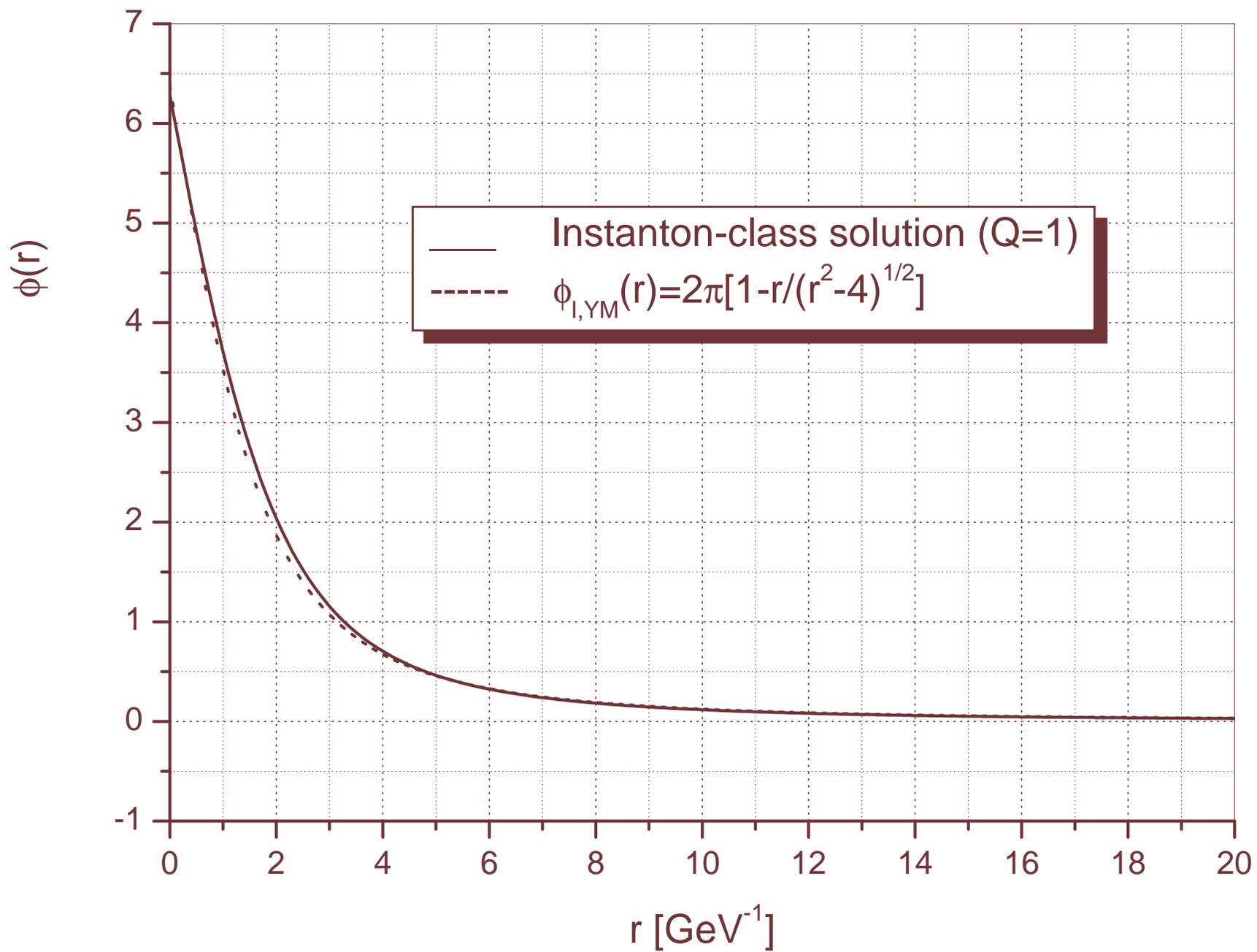
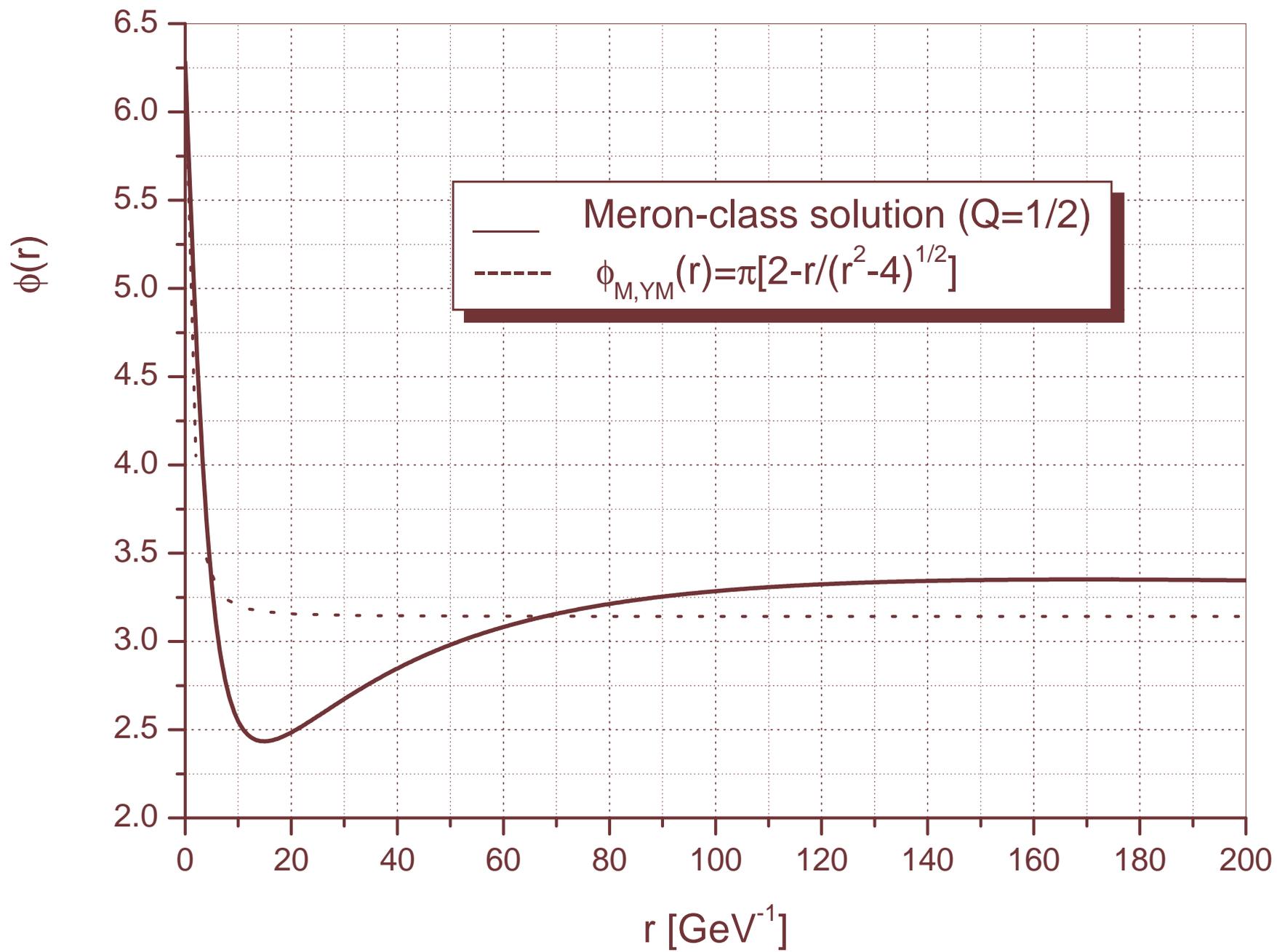
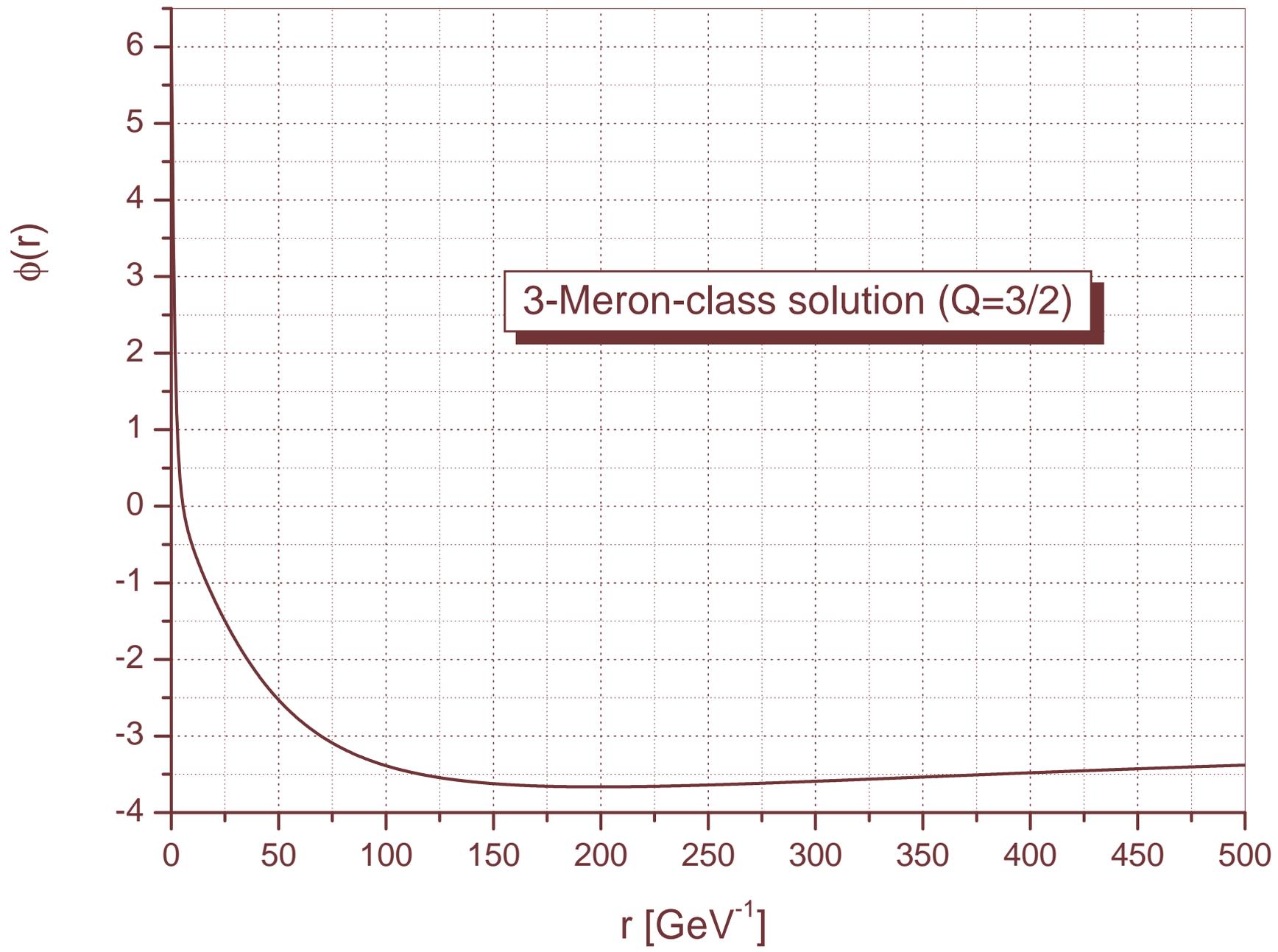
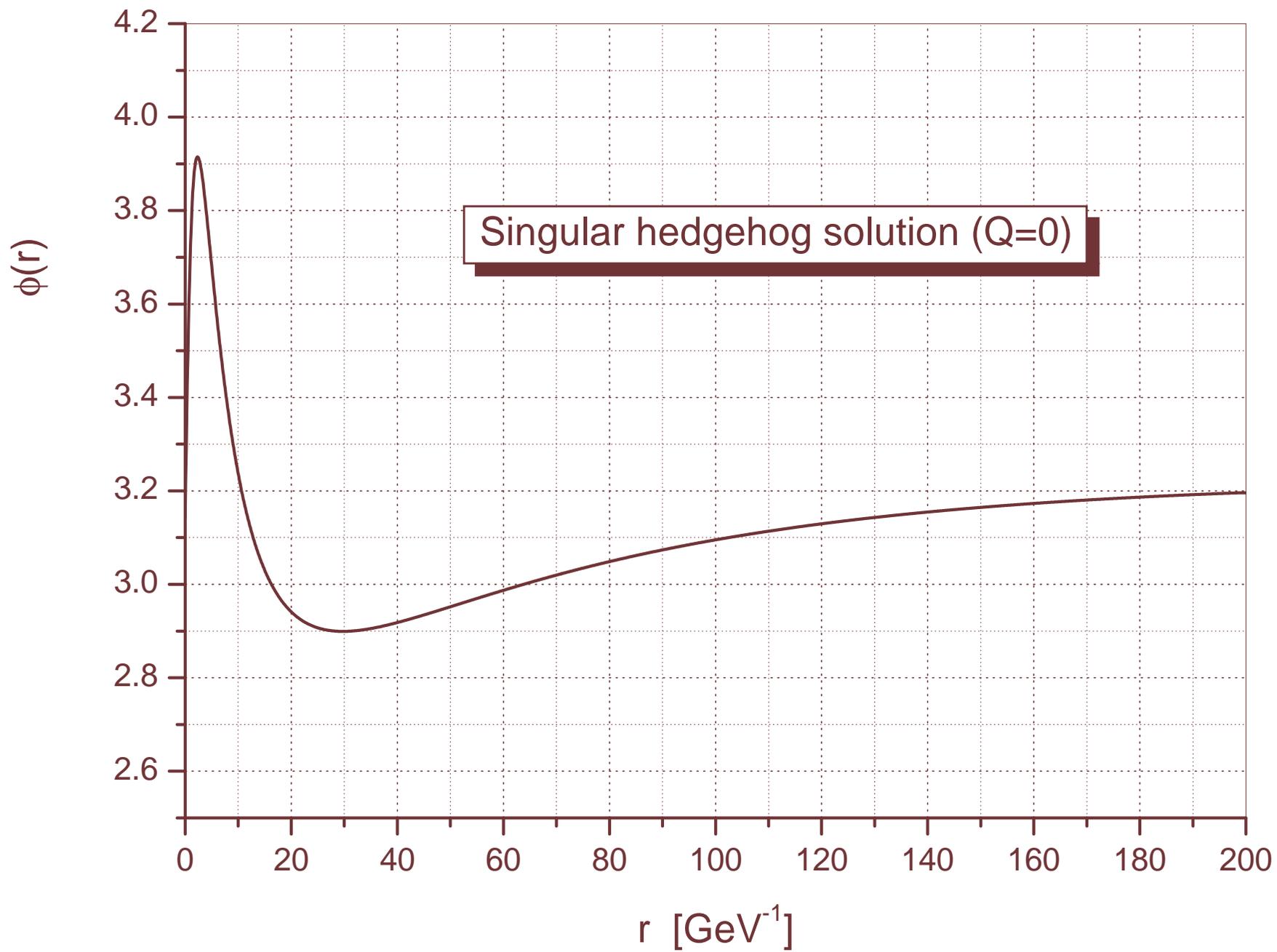
\end{document}